\documentclass[preprint]{aastex}
\usepackage{epsfig}

\begin{document}
\def\lsim{\raise0.3ex\hbox{$<$}\kern-0.75em{\lower0.65ex\hbox{$\sim$}}}
\def\gsim{\raise0.3ex\hbox{$>$}\kern-0.75em{\lower0.65ex\hbox{$\sim$}}}


\title{Does Magnetic Levitation or Suspension Define the Masses of 
Forming Stars?}

\author{Frank H. Shu}
\affil{National Tsing Hua University, 101, Section 2 Kuang Fu Road, 
Hsinchu, Taiwan 300, R.O.C.}

\author{Zhi-Yun Li}
\affil{Department of Astronomy, University of Virginia, Charlottesville, 
VA 22903}

\author{Anthony Allen}
\affil{Institute of Astronomy and Astrophysics, Academia Sinica, Taipei 106, 
Taiwan, R.O.C.}

\begin{abstract}
We investigate whether magnetic tension can define the
masses of forming stars by holding up the subcritical
envelope of a molecular cloud that
suffers gravitational collapse of its supercritical core.  We perform
an equilibrium analysis of the initial and final states assuming perfect 
field freezing, no rotation, isothermality, and a completely flattened 
configuration.
The sheet geometry allows us to separate the magnetic tension into a
qlevitation associated with the split monopole formed
by the trapped flux of the central star and a suspension
associated with curved field lines that thread the static pseudodisk
and envelope of material external to the star. We find solutions 
where the eigenvalue
for the stellar mass is a fixed multiple of
the initial core mass of the cloud.  We verify the analytically
derived result by an
explicit numerical simulation of a closely related 3-D axisymmetric
system.   However, with field freezing, the implied
surface magnetic fields much exceed measured values for young stars.  If
the pinch by the
central split monopole were to be eliminated by magnetic reconnection,
then magnetic suspension alone cannot keep the
subcritical envelope (i.e., the entire model cloud) from falling onto 
the star.
We argue that this answer has general validity, 
even if the initial state lacked any kind of symmetry, possessed rotation, and
had a substantial level of turbulence.
These findings strongly support a picture for the halt of infall that 
invokes dynamic
levitation by YSO winds and jets, but the breakdown of ideal 
magnetohydrodynamics
is required to allow the appearance in the problem of a rapidly rotating,
centrifugally supported disk.  We use these results to calculate
the initial mass function and star formation efficiency
for the distributed and clustered modes of star formation.
\end{abstract}


\pagebreak
\section{Introduction}

Despite the empirical evidence for the formation of stars under
a wide variety of physical conditions in the cosmos, and despite more
than a half-century of theoretical study, the
question of what determines the masses of forming stars from gravitational
collapse in large interstellar clouds remains an open one.
Ideas range from hierarchical, opacity-limited, thermal fragmentation (Hoyle 1953, Lynden-Bell 1973,
Rees 1976, Silk 1977, Bodenheimer 1978, Zinnecker 1984),
to magnetically-limited fragmentation (Mestel 1965, 1985),
to turbulence-induced fragmentation (Scalo 1985,1990; Larson 1995;
Elmegreen \& Efremov 1997; Truelove et al. 1998; Elmegreen 2002; Falgarone 2002;
Klein et al. 2003), to wind-limited mass infall (Shu \& Terebey 1984,
Shu et al. 2000).

Shu (1977) pointed out the intrinsic difficulty of
defining starlike masses in a large, unmagnetized, isothermal cloud that starts in
equilibrium or near-equilibrium and is not bounded by an artificial
surface pressure.  He commented on the difficulty of fragmentation in
such an environment, where the condition of equilibrium of the initial state and
the steep velocity gradients of the subsequent inside-out collapse effectively
prevent gravitational fragmentation of the type originally contemplated by
Hoyle from taking place (see also the debate between Hunter 1967 and Layzer 1964).
These comments were subsequently given added force in numerous numerical simulations
(see, e.g., Tohline 1980, 1981; Truelove et al. 1998).

Mestel (1965, 1985; see also Nakano \& Nakamura 1978)
made a crucial distinction between clouds that
have dimensionless mass-to-flux ratios greater than unity (supercritical -- capable
of gravitational collapse), and less than unity (subcritical -- incapable of gravitational
collapse).  He argued that spherical magnetized clouds which are marginally supercritical
cannot fragment gravitationally because any spherical subpiece will
be subcritical.  He speculated, however, that supercritical isothermal clouds
that collapse coherently to highly flattened states can
fragment into supercritical pieces, each with size comparable to the vertical
scale height.   These pieces might correspond
to stellar masses.  Shu \& Li (1997) cast doubt on this
scenario in the case when field freezing may be assumed and the 
initial state corresponds again to a state of initial force balance
with a significant density and field stratification.
Self-similar, inside-out, collapses in such circumstances do indeed show
the anticipated flattening, but the dynamically infalling pseudo-disks that
result exhibit no tendency to fragment gravitationally.  They
are prone to other kinds of numerical and perhaps physical instabilities
in the presence of non-ideal magnetohydrodynamic (MHD) effects
(Allen, Shu, \& Li 2003; 
Allen, Li, \& Shu 2003).  Indeed, to our knowledge, no reliable 
numerical simulation of initial non-uniform states of isothermal 
equilibria, with or without magnetization, has ever been able to 
demonstrate fragmentation in the subsequent collapse.
\footnote{If the cloud mass is higher than the equilibrium 
value and starts with a nearly homogeneous density distribution,
so as to contain either initially or shortly thereafter more than one Jeans mass,
then fragmentation is possible, as shown, for example, by Bodenheimer et
al. (2000) 
or Matsumoto \& Hanawa (2003).  But such initial states require explication
how they arose, since their evolutionary time scales are generally short in
comparison with the probable ages of the parent molecular clouds.}

Truelove et al. (1998) and
Klein et al. (2003) find fragmentation possible in unmagnetized
clouds that start with substantial supersonic turbulence.  However,
the entire self-gravitating cloud cannot resist being turned into stars in
the presence of strong shock dissipation (see, e.g., Klessen 2001).
This then introduces a problem why such clouds are around today in a
universe whose age is many dynamical crossings of molecular clouds
(Zuckerman \& Evans 1974).  Scenarios have been
proposed where the star formation occurring in molecular clouds
on all scales occupies only one dynamical
crossing time before being dispersed (Elmegreen 2000, Hartmann 2003).
But given the known overall inefficiency of star formation, these
proposals then need to elucidate how the small fraction of cloud matter
that came to have star-formation capability arrived at the critical
state, and how the vast bulk of the molecular cloud material avoids
gravitational collapse and fragmentation.\footnote{Lada \& Lada (2003)
estimate that less than 10\% of the mass of a giant molecular cloud
participates in star formation, and the star formation efficiency in
even the densest regions rarely exceeds 10--30\%.}

The problem of rapid turbulent dissipation
persists even in the presence of magnetization (Heitsch, MacLow,
\& Klessen 2001; Ostriker, Stone, \& Gammie 2001),
unless the magnetization is so strong as to
make the clouds subcritical, in which case neither runaway gravitational
collapse nor gravitational fragmentation will occur, even if the
turbulence dissipates completely.  In any case, in
prototypical low-mass cores of molecular clouds that form sunlike stars,
the level of turbulence is low (Myers 1995, Evans 1999), too subsonic 
to satisfy the supersonic conditions found necessary by 
Truelove et al. (1998) and Klein et al. (2003) for turbulent fragmentation.

The theoretical solution for making the vast bulk
of molecular cloud material have a low efficiency for star formation is
therefore simple, at least in principle: 
assume that the bulk of molecular clouds,i.e., their envelopes, are
subcritical (Shu, Adams, \& Lizano 1987).
The trick to gravitational collapse (and fragmentation)
is then to invent mechanisms to get pieces (the cores) that are supercritical.
Lizano \& Shu (1989, see also Shu et al. 1999) outlined a bimodal process to
accomplish this task: with ambipolar diffusion producing small, quiescent, supercritical
cores in the isolated (or distributed) mode of star formation, and cloud-cloud collisions
along field lines producing large, turbulent, supercritical cores in the cluster mode
of star formation.  Molecular-cloud turbulence could also play a role in accelerating the
rate of star formation in the distributed mode, both in concentrating matter
during the dissipation of turbulence (Myers \& Lazarian 1998, Myers 1999)
and in enhancing the effective speed of ambipolar diffusion by the effects of fluctuations
(Fatuzzo \& Adams 2002, Zweibel 2002).
We postpone further discussion of such
effects until \S 4.

Mouschovias (1976) made the intriguing suggestion that magnetic tension 
in a magnetized cloud might be able to hold up the envelope, preventing 
it from joining the mass of the cloud core that collapses to the 
center to form the central star. Despite initial enthusiasm for
this idea to define stellar masses (e.g., Shu 1977),
subsequent collapse calculations of 
supercritical clouds (or the parts of them that are supercritical)
showed that the original proposal based on curvature arguments
was not well-founded (Galli \& Shu 1993a,b;
Li \& Shu 1997; Allen et al. 2003a,b).  However, a modified form of
the question remains open for models where the cloud cores are 
supercritical but the envelopes are subcritical.

\subsection{Goal of this Work}

The present paper is motivated by the desire to settle the last issue.  
We set the discussion in the context of a very specific theoretical 
model, but we shall argue later that the answers derived are generic and 
suggestive.  Although we start with the picture of a cloud as a well-defined
entity, we actually have considerable sympathy for the
view of the turbulence camp that such a concept
has limited utility above the mass scales
of molecular cloud cores (see, e.g., Elmegreen 1995 or Larson 1995).
However, the discussion becomes more concrete if we save the debate
on the role of interstellar turbulence until the end of the paper.
 
In our formal calculations, we are interested in initial mass 
distributions given by the singular isothermal sphere (SIS),
\begin{equation}
\rho(r) = {a^2\over 2\pi G r^2} , 
\label{e-10}
\end{equation} 
where $a$ is the isothermal speed of sound in cosmic molecular gas,
$G$ is the universal gravitational constant, and $r$ is the radial
distance from the cloud center.
Like Galli \& Shu (1993a,b), we thread this cloud with a uniform
magnetic field of strength $B_0$ in the $z$ direction.  Since the field 
is uniform,
it exerts no MHD forces, and the force balance between self-gravity and
isothermal pressure represented by the (unstable) equilibrium (\ref{e-10}) 
is undisturbed.

In cylindrical coordinates $(\varpi, \varphi, z)$, the surface 
density corresponding
to a vertical projection of the SIS volume density (\ref{e-10}) 
along field lines
to the equatorial plane $z=0$ is given by
\begin{equation}
\Sigma (\varpi) \equiv \int_{-\infty}^{\infty} 
{a^2\, dz\over 2\pi G(\varpi^2+z^2)}= {a^2\over 2G\varpi} . 
\label{e-9}
\end{equation}
The corresponding mass enclosed in an infinite cylinder of radius $\varpi$ is
then
\begin{equation}
M(\varpi) = \int_0^\varpi \Sigma(\varpi) \, 2\pi \varpi d\varpi =
{\pi a^2 \over G} \varpi. 
\label{e-8}
\end{equation} 
The magnetic flux contained in the same cylinder is
\begin{equation}
\Phi(\varpi) = B_0 \pi \varpi^2 . 
\label{e-7}
\end{equation}
Thus, the differential mass-to-flux ratio reads
\begin{equation}
{dM\over d\Phi} = {a^2 \over 2 GB_0 \varpi} = 
{\pi a^4\over 2G^2 B_0 M}. 
\label{e-6}
\end{equation}
Notice that the surface density $\Sigma$ 
(projected mass per unit area) and the magnetic
field $B_0$ (flux per unit area)  have the same ratio as the 
differential mass to flux, $dM/d\Phi$. 

Following Basu \& Mouschovias (1994) and Shu \& Li (1997), we 
nondimensionalize
using $(2\pi G^{1/2})^{-1}$ as the basic unit of mass-to-flux
(see also Nakano \& Nakamura 1978).  Thus,
the dimensionless differential mass-to-flux ratio is given by
\begin{equation}
\lambda (\varpi) \equiv 2\pi G^{1/2} {dM\over d\Phi} =
{M_0\over M}, 
\label{e-5}
\end{equation}
where
\begin{equation}
M_0 \equiv {\pi^2 a^4\over G^{3/2} B_0} , 
\label{e-4}
\end{equation} 
is the fundamental mass scale in the problem.
We denote the cylindrical radius $\varpi$ corresponding to the critical
value $\lambda(\varpi) = 1$ by $r_0$; it is computed from equation 
(\ref{e-8}) when $M=M_0$ and $\varpi=r_0$:
\begin{equation}
r_0 = {\pi a^2\over G^{1/2} B_0} . 
\label{e-3}
\end{equation}
Because $M(\varpi)$ scales directly as $\varpi$ in our initial state,
we may also equivalently write equation (\ref{e-5}) as
\begin{equation}
\lambda (\varpi) = {r_0\over \varpi}. 
\label{e-2}
\end{equation}
We heuristically refer to the regions with $\varpi < r_0$,
where the initial cloud is supercritical, $\lambda > 1$,
as the ``cloud core;'' and the regions with $\varpi > r_0$,
where the cloud is subcritical, $\lambda < 1$,
as the``cloud envelope.''   

For $a=0.2$ km/s and $B_0=30 \; \mu$G believed to be typical
of dense regions in cold molecular clouds, we have
$r_0 = 0.05$ pc and $M_0= 1.5\; M_\odot$.
These are suggestive values
for the cores in the Taurus molecular cloud (Jijina, Myers, \& Adams
1999, Evans 1999), as has been noted in other contexts by many authors.
Low-mass cores in the most crowded regions of clustered star formation
might begin to overlap, unless such highly pressured regions have, as
likely, larger ambient values of $B_0$.
Alternatively, the whole region of embedded cluster formation
may be supercritical, and the magnetic difference
between core and (common) envelope loses some of its distinction (see \S 4.5).
In any case, the mass scale (\ref{e-4}) resembles a kind of magnetic Bonnor-Ebert mass
(Ebert 1955, Bonnor 1956), with $B_0^2/8\pi$ replacing the role of the
external pressure to define a
critical condition for gravitational collapse
(see the discussion of Shu et al. 1999 concerning why
such a concept only holds for the supercritical cores).
But the analogy is not completely apt, for reasons to be made clear
in this paper.

Galli \& Shu (1993 a,b) showed that an initial state with frozen-in 
magnetic fields corresponding to the mass-to-flux distribution 
$dM/d\Phi$ of the above configuration is
unstable to inside-out collapse, with the formation of a point object
at the center that grows in mass with time.  The corresponding
flux trapped at the center fans out as a split monopole eventually to
connect at large distances onto the straight and uniform field lines of
the initial state.  Galli \& Shu followed only the
initial stages of the inside-out collapse involving the supercritical regions
of the cloud, where the mass accumulation rate at the center has its
SIS value, $\dot M = 0.975 \, a^3/G$ (Shu 1977), even though the
infall now takes place through a pseudodisk rather than spherically
symmetrically onto the center.   

We wish to extend the problem considered by Galli
\& Shu (1993a,b) by asking the following questions. What is the
``final'' state of the collapse?  How much of the ``core'' or envelope
will eventually end up inside the star?  If it is not the
entire cloud, what is the multiple $m_*$ of the original supercritical core
$M_0$ that becomes stellar material?  By what mechanism
would the material beyond the amount $m_*M_0$ be prevented from falling 
into the central
star? How is the transition from the initial state to final
state made in time?  How sensitive are our answers to the assumption
of perfect field freezing, i.e., to the preservation of the function
$dM/d\Phi$ from beginning to end?

\subsection{Findings of this Paper}

We approach these questions by two kinds of calculations.  One is by
numerical simulation of the time-dependent evolution.  This simulation is
presented in \S 3. The other approach is
to attack directly the final-state equilibrium, with the value
of $m_*$ to be obtained as an eigenvalue of the problem.  The mathematical
formulation of the resultant problem as an integro-differential equation
in a single variable and its numerical solution occupy \S 2 of this
paper.  Readers uninterested in this mathematical derivation or its counterpart in
numerical simulation (\S 3) may jump directly to \S 4 from the end of \S 1.

The direct attack for the final state is extremely informative
and will be discussed first.
The full 3-D (axisymmetric) problem of magnetostatic equilibrium
with prescribed $dM/d\Phi$ is 
an involved and difficult calculation
(see Mouschovias 1976, Nakano 1979, Lizano \& Shu 1989, Tomisaka, Ikeuchi, \&
Nakamura 1989),
even without the complication
of an eigenvalue search for a central point mass.
We therefore attack a simplified
version of the problem posed above.  This simplification is motivated
by the expectation that the inner regions of the suspended cloud envelope
will be highly flattened by the same anisotropic magnetic forces that give
rise to the pseudodisks of the dynamical collapse calculations.  By adopting
a gas pressure that acts only in the horizontal directions rather than
isotropically in all three directions,
even the initial state flattens completely
(see below).  The calculation of magnetic forces
simplify considerably in a completely flattened geometry 
(Basu \& Mouschovias 1994).
In particular, only tension forces remain, and they can be computed from the flattened
distribution of currents that act as the source of the magnetic fields
as ``action at a distance'' (Shu \& Li 1997).  

As an additional bonus, Newtonian gravity
has the fortunate coincidence that the radial gravity of a 
singular isothermal
sphere is identical to a completely-flattened 
singular isothermal disk (SID) if they have the
same column density distributions.
In other words, the self-gravity of a
completely flattened distribution,
\begin{equation}
\rho(\varpi, z) = \Sigma(\varpi) \delta (z), 
\label{e-1}
\end{equation}
when $\Sigma(\varpi)$ is given by equation (\ref{e-9}), can be exactly
offset by the isothermal (2-D axisymmetric), negative gas-pressure gradient,
$-a^2 d\Sigma/d\varpi$, of this SID. 
Thus, the SID can also be threaded by an initially uniform, vertical
magnetic field of strength $B_0$ and remain in (unstable) equilibrium.  Such
a SID is no longer magnetized {\it isopedically} in the nomenclature
of Li \& Shu (1997), and it will undergo inside-out gravitational
collapse in a non-self-similar manner.  Nevertheless, we 
shall make good use of the remaining correspondences between the 
non-self-similar, axisymmetric, 2-D and 3-D problems in what follows.
In particular, written in terms of the variables defined previously,
the local value of the dimensionless
mass-to-flux ratio of the completely flattened configuration,
\begin{equation}
{2\pi G^{1/2}\Sigma(\varpi)\over B_0} = \left( {r_0 \over \varpi}\right),
\label{e0}
\end{equation}
recovers our identification that $\varpi = r_0$ marks the cylindrical radius
where our model SID makes a transition from being supercritical
(cloud core) to subcritical (cloud envelope).  For $B_0 = 30 \; \mu$G,
the transition between core and envelope is made at a surface density
equal to $B_0/2\pi G^{1/2} = 1.8\times 10^{-2}$ g cm$^{-2}$, which corresponds
to about 4 magnitudes of visual extinction, a suggestive value from the
points of view of observations (Blitz \&  Williams 1999) and
magnetically limited star formation (McKee 1989).

The gravity and magnetic forces will be different for the intermediate regions
in the time-dependent collapse of the axisymmetric 2-D and 3-D problems.  Near the origin,
the gravity and magnetic forces are dominated by the central object, which is
a point mass and a split monopole in both calculations.
Because the pseudodisk
is very thin in its innermost regions even for the formal 3-D problem, it
may not be a surprise to find that the eigenvalue $m_*$ for the final state,
assuming field freezing,
is the same for both problems: $m_* = \surd 2$ (see below).  
We should add the immediate caveat that the answer $m_* = \surd 2$ is an {\it
exact} result of the 2-D axisymmetric calculation, whereas it is less precisely
determined in the 3-D calculation.  But because the analysis shows that
this result depends only on what happens close to the central source,
and not at all on what happens in the outer envelope, we have reason to believe
that the result holds accurately for both configurations.  

The mechanism of holding up the envelope is counter-intuitive,
as hinted upon by the description given just now
that the eigenvalue $m_*$ for the mass of the central object is 
determined entirely
by what happens near the star. 
Instead of the envelope being suspended by magnetic tension working
against the gravity of the cloud core and star, 
the envelope is being levitated by the split monopole pushing against the
background magnetic field in the vacuum regions above and below
the pseudodisk.  

However, in order to hold off the inflow by this process,
the central star would need to trap fields 
of $\sim 10$ megagauss.
This value is a factor of about 5000 times larger than the fields 
measured in low-mass pre-main-sequence stars
(Basri, Marcy, \& Valenti 1992; Johns-Krull,
Valenti, \& Koresko 1999).
Evidently, through flux leakage to the surroundings, or
through flux destruction via magnetic reconnection or anti-dynamo action,
young stars destroy much if not all of the interstellar fields
brought in by the mass infall.  In \S 4, we shall comment
on the implications for the overall problem by these various possibilities.

The tentative answer to the
question posed in the title of this paper is, therefore,
an equivocal ``no.''  The equivocation arises because the time-dependent
simulations show, as was intuitively expected before we began the actual
calculations, a significant reduction in the rate $\dot M$ of mass infall
onto the central source as the outwardly propagating wave of infall 
expands into the subcritical envelope.  Although the rate can be reduced 
all the way to zero only with the help of the magnetic push extended by 
the central split monopole,
the mechanism for the reduction resides partially in the
strong fields (relative to gas pressure) in the cloud envelope and not
only in the magnetic levitation provided by the central source.  The
two effects interact in a complex fashion to produce the final result.
Indeed, before the paper ends, we will have revised our answer to
a qualified ``yes!''

In this regard,
the reader should not be fooled by the usual expression for magnetic tension,
$({\bf B}\cdot \nabla){\bf B}/4\pi$, into thinking that this force has
a purely local origin.  The local expression is useful only if we have 
other means
to compute the field ${\bf B}$ and its derivative along field lines. 
Those other means, i.e., the equations
of magnetohydrodynamics, include effects that mimic ``action at a
distance,'' particularly in the near-vacuum conditions above and below
the pseudodisk and in the outer cloud envelope where Alfv\'en waves
can propagate almost with infinite speed relative to $a$. This 
``action at a distance''
brings magnetic influences from throughout the system, including the 
origin and the outermost regions.  Although we reject magnetic
levitation by a trapped split monopole at the center as a viable
practical mechanism in the long run, its role in halting
infall can be effectively
replaced by the magnetized winds and jets seen in actual protostars
(Lada 1985, Bachiller 1996, Reipurth \& Bally 2001, Shang et al. 2002).
These are believed to
arise because the realistic problem includes rotation in addition to
magnetic fields (K\"onigl \& Pudritz 2000, Shu et al. 2000).
Even if the winds shoot purely out of the plane
of an idealized 2-D axisymmetric calculation, they would still have an effect
in blowing away the gas in a thin disk or pseudodisk because of their
action on the fields from above and below that thread the flattened
distribution of matter.  We defer this discussion, however, to the
concluding remarks at the end of this paper.

\section{Geometrically-Flat Final State}

\subsection{Derivation of Basic Forces in a Completely Flattened Geometry}

We begin our analysis by examining the final state of the 2-D axisymmetric problem posed
in \S 1.
In the approximation that we treat the gas pressure-tensor as having no
vertical component, the entire matter and current
configuration is confined to a sheet in the plane $z=0$.  In the notation of
equation (2.4) of Shu \& Li (1997), then,
the tension force per unit area acting on the sheet is
\begin{equation}
{B_z{\bf B}_\parallel^+ \over 2\pi}, 
\label{e1}
\end{equation} 
where ${\bf B}_\parallel^+$ is the horizontal component
of the magnetic field at the upper surface of the sheet.
Since ${\bf B}_\parallel$ reverses directions below the
sheet, whereas $B_z$ remains continuous in magnitude and direction
upon crossing the sheet, the total field ${\bf B}={\bf B}_\parallel +
B_z \hat {\bf e}_z$ has a kink at the midplane.  Through Amp\'ere's law
this kink is supported by an electric sheet-current ${\bf J}$.
The Lorentz force per unit area, given by the cross product of
${\bf J}$ and the magnetic field in the midplane $B_z\hat {\bf e}_z$ divided
by the speed of light,
yields the tension force per unit area displayed in 
equation (\ref{e1}).  The dragging of
the inner portions of this matter and field configuration into the
origin by gravitational collapse produces the point mass and
split monopole that play such central roles in our formal analysis
of the final state.  

To begin in a general way, we shall assume arbitrary variations in space and time
for the flattened geometry.  Above the sheet exists a vacuum,
and the current-free (curl-free)
magnetic field is derivable from a scalar potential.  Indeed, since
a uniform field $B_0\hat {\bf e}_z$ is also a vacuum field, we may
first subtract off such a uniform field from the total ${\bf B}$,
\begin{equation}
\hat {\bf B} \equiv {\bf B}-B_0\hat {\bf e}_z , 
\label{e2}
\end{equation}
and derive $\hat {\bf B}$ from a scalar potential $\Psi$,
\begin{equation}
\hat {\bf B} = \nabla \Psi . 
\label{e3}
\end{equation}
This field also has zero divergence, so $\Psi$ satisfies
Laplace's equation,
\begin{equation}
\nabla^2 \Psi = 0 \qquad {\rm for} \qquad z > 0. 
\label{e4}
\end{equation}
The boundary condition (2.7) of Shu \& Li (1997) is modified to read
\begin{equation}
{\partial \Psi\over \partial z} = B_z-B_0 \qquad {\rm on}
\qquad z = 0^+; 
\label{e5}
\end{equation}
whereas the boundary condition (2.8) reads as before:
\begin{equation}
\hat {\bf e}_r\cdot \nabla \Psi \rightarrow 0 \qquad {\rm as}
\qquad r = (\varpi^2+z^2)^{1/2} \rightarrow \infty. 
\label{e6}
\end{equation}
Comparison of equations (\ref{e4})--(\ref{e6}) with the gravitational 
potential
problem of thin disks shows that the solution for $\Psi(\varpi,\varphi, z, t)$
is given by the Poisson integral:
\begin{equation}
\Psi (\varpi,\varphi, z, t) = -{1\over 2\pi}\int_0^\infty r\, dr
\oint {[B_z(r,\psi, t)-B_0] \, d\psi\over [\varpi^2+r^2-2\varpi r\cos(\varphi-\psi)+z^2]^{1/2}}, 
\label{e7}
\end{equation}
with $B_z$ evaluated in the plane $z=0$. 
The tension force acting per unit area of the pseudodisk
in the horizontal directions is now given by
\begin{equation}
B_z(\varpi,\varphi, t) \int_0^\infty r \, dr
\oint [B_z(r,\psi,t)-B_0]{\hat e_\varpi [(\varpi-r\cos(\varphi-\psi)]+
\hat e_\varphi r\sin(\varphi-\psi)\over (2\pi)^2 [\varpi^2+r^2-2\varpi r\cos
(\varphi-\psi)]^{3/2}} \, d\psi. 
\label{e8}
\end{equation} 

Notice that the interaction of the magnetic fields
in the vacuum regions above and below the electrically
conducting sheet delocalizes the instantaneous tension force felt by the sheet.
The strength of the magnetic field
relative to its unperturbed value in the sheet,
$B_z(r, \psi, t)-B_0$, at a footpoint location $(r, \psi)$, acts as a source
for exerting tension force at the field point $(\varpi, \varphi)$,
all at time $t$, by ``action at a distance.''

\subsection{Force Balance in a Flat Pseudodisk 
Surrounding a Magnetized Point Mass}

Using an analogous expression for the gravitational force acting
on the pseudodisk per unit area, and adopting axial symmetry and
time-independence, the azimuthal components of magnetic and
gravitational force vanish by
symmetry, and we may write the condition of radial
balance of axisymmetric 2-D gas pressure, magnetic tension,
and gravity as (cf. Shu \& Li 1997):
\begin{equation}
-{d\Pi\over d\varpi}+{1\over \varpi^2}
\int_0^\infty K_0\left({r\over \varpi}\right)
\left[{B_z(\varpi)[B_z(r)-B_0]\over (2\pi)^2}-
G\Sigma(\varpi) \Sigma(r) \right] \, 2\pi r  dr = 0 , 
\label{e9a}
\end{equation}
where $\Pi$ is the usual gas pressure integrated over the disk thickness:
\begin{equation}
\Pi = a^2\Sigma, 
\label{e9b}
\end{equation}
and $a^2=kT/\mu$ where $k$ is Boltzmann's constant, $\mu$ is the mean molecular
mass, and $T$ is the local gas temperature (which can be made
a function of $\varpi$ for the final state
if we wish to add additional realism). 

In equation (\ref{e9a}) $K_0(r/\varpi)$ is the normalized radial-gravity kernel for
axisymmetric thin disks,
\begin{equation}
K_0(\eta) \equiv {1\over 2\pi}\oint {(1-\eta
\cos \varphi)\, d\varphi\over (1+\eta^2-2\eta 
\cos \varphi)^{3/2}} . 
\label{e10}
\end{equation}
Note that $K_0(0) = 1$ and $K_0(\eta) \rightarrow -(1/2)\eta^{-3}$ for
$\eta \rightarrow \infty$.  Figure 1 displays $K_0(\eta)$ and its
associated ``potential'' function $H(\xi)$ with $\xi = 1/\eta$
(see Appendix A).

We suppose that the origin contains a point mass and a split monopole,
so that the integrations of $\Sigma(r)$ and $B_z(r)$ over $r$
include delta functions at the origin $r=0$.  Splitting off these terms
explicitly, and making use of the property $K_0(0) = 1$, we obtain
\begin{equation}
-{\varpi^2 \over G\Sigma(\varpi)}{d\Pi \over d\varpi}+
\left[ {1\over \lambda_* \lambda(\varpi)}-1\right]M_*
+ \int_{0^+}^\infty K_0\left({r\over \varpi}\right)
\left\{ {1\over \lambda(\varpi)}\left[{1\over
\lambda(r)}-{1\over \lambda_0(r)}\right]-1\right\}
\Sigma(r) \, 2\pi r dr = 0, 
\label{e11}
\end{equation}
where we have denoted
\begin{equation}
\lambda(\varpi) \equiv 2\pi G^{1/2}{\Sigma(\varpi)\over B_z(\varpi)},
\label{e12}
\end{equation} 
\begin{equation}
\lambda(r) \equiv 2\pi G^{1/2}{\Sigma(r)\over B_z(r)},
\label{e13a}
\end{equation}
\begin{equation}
\lambda_0(r) \equiv 2\pi G^{1/2}{\Sigma(r)\over B_0}
= {B_z\over B_0}\lambda(r),
\label{e13b}
\end{equation}
as, respectively, the dimensionless mass-to-flux
ratio in the pseudodisk of the field point, the source point,
and the correction for the background field.  By $\lambda_*$
we mean the average value of the mass-to-flux ratio
of the central point:
\begin{equation}
\lambda_* \equiv 2\pi G^{1/2}{M_*\over \Phi_*} . 
\label{e14}
\end{equation}

The division of the original integral into two parts requires us to introduce 
nomenclature to avoid possible confusion.  The magnetic tension force
at field point $\varpi$ is the integral of all sources from $0$ to $\infty$
in equation (\ref{e9a}), which includes the
point source at the origin.  We call the separate forces arising
from the split monopole at the origin and from the integral
from $0^+$ to $\infty$ as, respectively, ``magnetic levitation'' 
and ``magnetic suspension.''
At the most picturesque level, levitation is an influence that comes 
from below (near or at the position of the star) whereas suspension
comes from above (near or beyond the outer envelope). 
The basic contention of this paper is that mass infall can be
halted by levitation, but not by suspension.  An explicit proof
follows for a specific example; a more general argument is developed
in \S 4.

\subsection{Masses and Fluxes Assuming Field Freezing}

So far, our formulation has been quite general, apart from the
assumptions of axial symmetry, force balance, and a 
completely flattened geometry. 
We now specialize by adding the following assumptions: (a) isothermality,
i.e., $a^2$ = same constant for initial and final states,
and (b) the conservation of the mass-to-flux distribution $dM/d\Phi$ of equation (\ref{e-5}).
We return in \S 4 to discuss the relaxation
of these additional assumptions.  For now, we merely note that (a) the
gas pressure in the pseudodisk is never a large term
relative to the others except in the outer envelope,
where $a$ in the final and initial states can be expected
to have similar values, and (b) changing the mass-to-flux distribution
for the final state corresponds to relaxing the constraint of
field freezing in the transition from initial to final state.

To implement the constraint of field freezing,
we find it convenient to use the running cylindrical mass
in place of the surface density:
\begin{equation}
dM = \Sigma (r) \, 2\pi r dr , 
\label{e16}
\end{equation}
with the convention that $M(r=0^+) = M_*$.
The corresponding flux variable is defined by
\begin{equation}
d\Phi = B_z(r) \, 2\pi r dr = 2\pi G^{1/2}{dM\over \lambda(r)}, 
\label{e17}
\end{equation}
with the similar convention that $\Phi(r=0^+) = \Phi_*$.
If we assume field freezing in the transformation from initial 
to final state, the differential mass-to-flux ratio is given
as the invariant function of equation (\ref{e-6}).

The integration of equation (\ref{e-6}) leads to the identification,
\begin{equation}
\Phi_* = \pi G^{1/2} {M_*^2\over M_0}, 
\label{e18}
\end{equation}
which yields from equation (\ref{e14}) the result,
\begin{equation}
\lambda_* = {2M_0\over M_*}, 
\label{e19}
\end{equation}
where $M_0$ is defined by equation ({\ref{e-4}).
Henceforth, we nondimensionalize using $M_0$ as the mass scale
and $r_0$ as defined by equation (\ref{e-3})
as the length scale. 
These choices turn out to best simplify the numerical 
coefficients in the final
governing equation (see the comment following eq. [\ref{e25}]).

\subsection{Nondimensionalization}

We now introduce the dimensionless radius
\begin{equation}
x \equiv {\varpi \over r_0} , 
\label{e23}
\end{equation}
and dimensionless masses,
\begin{equation}
m_* \equiv {M_*\over M_0}, \qquad m(x) \equiv {M(\varpi)\over M_0}, 
\label{e24}
\end{equation}
with analogous definitions using $y$ instead of $x$ when the source point $r$ replaces
the field point $\varpi$.
In these variables, the $\lambda$'s read
\begin{equation}
\lambda_* = {2\over m_*}, \;\; \lambda(\varpi) = {1\over m(x)},\;\;
\lambda_0(r) = {m^\prime(y)\over y},
\label{e25}
\end{equation}
where we have substituted equation (\ref{e17}) into equation
(\ref{e13b}) to express $\lambda_0(r)$,
used equations (\ref{e-3}) and (\ref{e-4}) to identify
a coefficient of $G^{1/2}M_0/B_0r_0^2$ as unity, and
denoted $dm/dy$ as $m^\prime(y)$.

The 2 instead of a 1 in
the relationship between $\lambda_*$ and $m_*$ is physically meaningful.
It arises because, in the build up to the final mass,
the central point accumulates material with a decreasing
mass-to-flux ratio.  With a $dM/d\Phi$ that decreases
as $M^{-1}$, starting at $M=0$, the average mass-to-flux
value is always twice as large as the last piece of matter
and trapped flux to enter the star.

Equation (\ref{e11}) now reads
\begin{equation}
-{x^3\over m^\prime (x)}{d\over dx}\left[{m^\prime (x)\over x}\right]
+m_*\left[ {m_*\over 2} m(x) -1\right]
+\int_{0^+}^\infty K_0\left({y\over x}\right)
\left\{ m(x) \left[ m(y)m^\prime(y) -y\right] -m^\prime(y)\right\} \, dy = 0 .
\label{e26}
\end{equation}
The variation of dimensionless mass with dimensionless
radius, $m(x)$ or $m(y)$, is the unknown
function that is to be determined 
from the solution of the integro-differential
equation (\ref{e26}), while $m(0^+)$ = $m_*$ = constant
is the eigenvalue of the problem.

\subsection
{Eigenvalue}

To find the eigenvalue $m_*$, consider the behavior of equation (\ref{e26})
as we approach the origin $x\rightarrow 0^+$.  In this limit,
\begin{equation}
K_0\left({y\over x}\right) \rightarrow -{x^3\over 2y^3}, 
\label{e27}
\end{equation}
for all $y>0$, except for a negligibly small interval
in the integration of equation (\ref{e26}) near the origin $y=0^+$.
The function $m(y)$ is well behaved; it starts with a value $m_*$
at $y=0^+$, and monotonically increases to $m(y)\rightarrow y$
as  $y\rightarrow \infty$.  The linear divergence of $m(y)$ at
large $y$ is not enough in equation (\ref{e26}) to offset the
$y^{-3}$ rate of vanishing of $K_0(y/x)$ in the integral,
which represents the force contributions of the pseudodisk proper (all forces
being multiplied by $x^2$).
Similarly, the first term in equation (\ref{e26}) representing the negative
(specific) pressure gradient is vanishingly small ($\propto x$)
if $m^\prime (x)$ is a constant as $x\rightarrow 0^+$.
Thus, the only nonvanishing terms on the left-hand side of
equation (\ref{e26}) in the small-radius limit
$x\rightarrow 0^+$ are, not surprisingly,
the magnetic and gravitational influences from the split monopole
and point mass at the center:
\begin{equation}
m_*\left[{m_*^2\over 2}-1\right] = 0. 
\label{e28a}
\end{equation}
The solutions for possible eigenvalues are therefore
\begin{equation}
m_* = 0, \qquad {\rm or} \qquad m_* = \surd 2. 
\label{e28b}
\end{equation}

The eigenvalue $m_*=0$ has an exact
eigenfunction solution.  This solution corresponds to
a pressure-supported, self-gravitating SID threaded by a force-free
uniform magnetic field:
\begin{equation}
m(x) = x \qquad {\rm for} \qquad m_* = 0. 
\label{e29a}
\end{equation}
To verify that equation (\ref{e29a}) constitutes an exact solution
for equation (\ref{e26}), note that the latter becomes, with $m_*=0$
and $m(x)=x$:
\begin{equation}
\int_{0^+}^\infty K_0\left( {y\over x}\right) \, dy = x, 
\label{e29b}
\end{equation}
which is a mathematical identity, because
\begin{equation}
\int_0^\infty K_0(\eta) \, d\eta = -\int_0^\infty H^\prime (\xi) \, d\xi
= H(0) = 1, 
\label{e30a}
\end{equation}
where $H(\xi)$ is defined by equation (\ref{e46}) of the comments in the 
Appendix A (see also Fig.~1). We henceforth refer to equation (\ref{e29a}) 
as the {\it initial state} of the configuration.

We imagine the sheet of magnetized gas represented by equation (\ref{e29a})
to collapse from inside-out, conserving mass-to-flux, to produce
the final-state configuration with $m_* = \surd 2$.
The final state represents not 
only the exact solution for the idealized problem posed in
this paragraph, but it also gives an accurate portrayal of the
state of affairs for $x<<1$ in our later 3-D collapse simulations.

The transition from initial state to final state represented by equation
(\ref{e28b}) has a simple interpretation.  A 
nondimensional central mass $m_* = \surd 2$ for the final state yields a dimensionless
stellar mass-to-flux ratio,
\begin{equation}
\lambda_* = {2\over m_*} = \surd 2, 
\label{e30b}
\end{equation}
that is only moderately supercritical, with supercriticality
being a necessary condition for
any self-gravitating object which is not bounded by external pressure.
On the other hand, the material that last entered the star had
a dimensionless mass-to-flux ratio,
\begin{equation}
\lambda (0^+) = {1\over m(0^+)} = {1\over m_*} = {1\over \surd 2},
\label{e31}
\end{equation}
that is only moderately subcritical.  Indeed the innermost part
of the pseudodisk hanging
precariously above the star's equatorial regions
has this value of subcriticality, $\lambda = 0.707$.  It is prevented
from dropping onto the star, not because of suspension
forces in the pseudodisk, but because the magnetic fields
threading through the pseudodisk are severely pushed back by the
huge split-monopole fields of the central star.  To sustain
a mass-to-flux ratio of $\lambda_* = \surd 2$, a low-mass protostar
would require, in practice, surface fields of $\sim 10^7$ gauss.
Such values are many thousands of times larger than measured for
T Tauri stars.  Aside from the observational implausibility
of such huge fields having been brought into the star in the first place
without slippage or annihilation, or having been retained in the inner parts
of the suspended pseudodisk against similar nonideal MHD effects,
one can question whether such an extraordinary feat of
magnetic levitation is physically stable against non-axisymmetric
(magnetic Rayleigh-Taylor) overturn.  We return to these physical issues
in \S 4.  For now,
we continue with the mathematical discussion of the posed idealized problem.

\subsection
{Eigenfunction}

For $m_*=\surd 2$, we choose to write equation (\ref{e26}) in the form,
\begin{equation}
{x^2 m^{\prime\prime}(x)\over m^\prime (x)} = x + m(x) \left\{ 
1+\int_{0^+}^\infty K_0\left( {y\over x}\right) \left[m(y)m^\prime(y)
-y\right]\, dy \right\} 
- \left[\surd 2 + \int_{0^+}^\infty K_0\left( 
{y\over x}\right) m^\prime(y)\, dy \right].
\label{e32}
\end{equation}
This equation is to be solved
for the eigenfunction $m(x)$ subject to the two-point boundary conditions (BCs):
\begin{equation}
m(x\rightarrow 0^+) = \surd 2; \qquad m(x\rightarrow \infty) = x. 
\label{e33}
\end{equation}

Superficially, equation (\ref{e32}) seems to imply trouble for force balance
in the outer envelope, where we expect $m(x)$ to approach its unperturbed
value $x$.  Substitution of this relation, if it were exact, results
in an inconsistency: $0 = x+x-\surd 2-x$, where the first $x$ comes from
the gas pressure, the second $x$ comes from the split monopole, the $-\surd 2$
comes from the point mass, and the $-x$ comes from the pseudodisk self-gravity
(all terms are to be divided by $x^2$ to obtain the dimensionless 
accelerations).  Clearly, the self-gravity of a disk with surface density
$\propto 1/x$ is able to balance only one of the positive terms, gas
pressure or split-monopole, and the negative pull of the point mass,
$-\surd 2$, is unable to compete with the left-over $x$ at large $x$. 
We shall see, however, that a relatively small adjustment from
the unperturbed state, $m(x)=x+\epsilon(x)$,
of the outer envelope is able to resolve this apparent contradiction.
Only small adjustments are needed because the background
magnetic field in the very subcritical
parts of the cloud envelope for $x \gg 1$  
is very strong relative to the gas pressure, and only a slight bending
by this field is sufficient to produce suspension forces that offset
the imbalanced parts of the gas pressure or split monopole.  Notice,
however, that these latter quantities contribute a net force that is
{\it outwards}.  Thus, the cloud envelope needs to make an adjustment that
produces an offsetting {\it inwards} force.  How the suspension contribution
can be {\it negative} will be elucidated below. 

From $m(x)$, we can recover the surface density in the pseudodisk:
\begin{equation}
\Sigma (\varpi) = {B_0\over 2\pi G^{1/2}} {m^\prime (x)\over x}.
\label{e34}
\end{equation}
The behavior at large $x$, $m(x) \rightarrow x$,
represents the surface density implied of the initial SID, which
has in turn the value given by 
vertical projection (i.e., along straight
field lines) of the SIS onto the equatorial plane.  Near the
star (values of $x\sim 10^{-5}$ for practical applications),
however, we expect the surface density of the final-state pseudodisk to
drop below the value given by the
simple inward extrapolation of the power law, $\Sigma \propto x^{-1}
\propto \varpi^{-1}$.  After all, some of the inner material has vanished
into the central star.  

\subsection{Numerics}

Let us transform the dependent variable:
\begin{equation}
m(x) \equiv x+\epsilon (x) . 
\label{e35}
\end{equation}
Without any approximation, equation (\ref{e32}) now becomes
\begin{equation}
{x^2 \epsilon^{\prime\prime}(x)\over 1+\epsilon^\prime (x)} =
  [x+\epsilon(x)][1+{\cal J}(x)]-[\surd 2+{\cal I}(x)], 
\label{e36a}
\end{equation}
where
\begin{equation}
{\cal I}(x)=\int_{0^+}^\infty K_0\left( 
{y\over x}\right) \epsilon^\prime(y)\, dy , 
\label{e36b}
\end{equation}
and
\begin{equation}
{\cal J}(x)= \int_{0^+}^\infty K_0\left( {y\over x}\right) 
\left[\epsilon(y)+y \epsilon^\prime(y)+\epsilon(y)\epsilon^\prime(y) 
\right]\, dy.
\label{e36c}
\end{equation}
Equation (\ref{e36a}) has the following physical interpretation.  The 
analytical
subtraction of the balanced pressure and self-gravitational forces
of the unperturbed initial state, using equation (\ref{e29b}), 
makes equation (\ref{e36a}) an exact requirement for the balance 
of the {\it nonlinear perturbations}
associated with the collapse to a final state of equilibrium.  The left-hand
side represents the acceleration associated with perturbational pressure;
the first term on the right-hand side, $\propto 1+{\cal J}(x)$, represents
the acceleration associated with the perturbational magnetic tension 
(including both the split monopole and the background field); and the
second term on the right-hand side $\propto -[\surd 2+{\cal I}(x)]$, 
represents
the acceleration associated with the perturbational gravity (including both
the mass point and the self-gravity of the pseudodisk). 

The mathematical advantage of defining the two functions, ${\cal I}(x)$
and ${\cal J}(x)$ will become obvious shortly.
We expect the new dependent variable $\epsilon(x)$ to be a 
bounded function. It equals $\surd 2$ at $x=0^+$ and monotonically
approaches zero as a positive value for large $x$.
Thus, it should be more amenable to accurate numerical solution than
its counterpart $m(x)$.

We solve equation (\ref{e36a}) by iteration as if it were a second-order ODE,
\begin{equation}
x^2 {d^2\epsilon \over dx^2} - {\cal R}(x){d\epsilon\over dx} = {\cal R}(x),
\label{e37}
\end{equation}
where ${\cal R}(x)$ is given by the right-hand side of 
equation (\ref{e36a}).  
Equation (\ref{e37}) is to be solved subject to the two-point BCs:
\begin{equation}
\epsilon(x) = \surd 2 \;\; {\rm for} \;\; x \rightarrow 0^+,
\qquad \epsilon (x) \rightarrow A_2x^{-2}
\;\; {\rm for} \;\; x\rightarrow \infty,
\label{e38}
\end{equation}
with $A_2$ to be determined as part of the numerical solution.

The requirement that $\epsilon(x) = A_2x^{-2}$
at large $x$ provides a good fit for the behavior
of the numerical solution and can be justified by an asymptotic analysis
(see Fig.~2 below and Appendix B).
Appendix B demonstrates that the
functions ${\cal I}(x)$ and ${\cal J}(x)$ have the
asymptotic properties:
\begin{equation}
{\cal I}(x) \rightarrow -\surd 2 \qquad {\rm as} \qquad x\rightarrow \infty,
\label{e39a}
\end{equation}
\begin{equation}
{\cal J}(x) \rightarrow -1 \qquad {\rm as} \qquad x\rightarrow \infty .
\label{e39b}
\end{equation}
These properties guarantee the good behavior of the 
right-hand side of equation (\ref{e36a}) in the limit of large $x$.
They also imply the physically interesting, individual rather than summed,
balance asymptotically of the perturbation self-gravity
of the pseudo-disk against the pull of the central mass point,
and of the perturbation suspension of the pseudodisk  
against the push of the central split monopole.  The mediation of 
the perturbation gas pressure in taking care of small residual
forces not balanced by these two large effects is compatible
with the declining
power law of the excess enclosed mass $\epsilon(x)\propto x^{-2}$
resulting from the enhanced gravitational pull
of the collapsed final state described in equation (\ref{e38}).

The two-point BCs (\ref{e38}) suggest that we attack the governing 
ODE by the Henyey
technique.  We introduce a mesh, $k=0,1,\dots,N, N+1$ such that 
$x_k = kh$, with
$h\ll 1$ and $N \gg 1$ in such a manner that $Nh$ is a moderately large
number.  We then discretize the governing ODE,
\begin{equation}
{x_i^2\over h^2}(\epsilon_{i+1}-2\epsilon_i+\epsilon_{i-1})
-{\cal R}_i {1\over 2h}(\epsilon_{i+1}-\epsilon_{i-1}) = {\cal R}_i, 
\qquad i = 1, \dots, N, 
\label{e40a}
\end{equation}
where 
\begin{equation}
{\cal R}_i = (x_i+\epsilon_i)(1+{\cal J}_i)
-(\surd 2+{\cal I}_i),
\qquad i =1, \dots, N .
\label{e40b}
\end{equation}
To evaluate ${\cal I}_i$ and ${\cal J}_i$, we note the total derivative nature
of the factors $\epsilon^\prime(y)$ and $\epsilon(y)+y \epsilon^\prime(y)+
\epsilon(y)\epsilon^\prime(y) = d[y \epsilon(y) + \epsilon^2(y)/2]
/dy$ that are the source terms in the integrands for ${\cal I}(x)$ and 
${\cal J}(x)$.  As a result, we can put the entire integrand halfway between
grid points and not have any integration weights (to second-order accuracy)
when we replace integrals by sums:
\begin{equation}
{\cal I}_i ={\cal C}_i+\sum_{k=0}^{N} K_0\left({y_{k+1/2}\over x_i}\right)
	(\epsilon_{k+1}-\epsilon_k),
\label{e41a}
\end{equation}
and
\begin{equation}
{\cal J}_i={\cal D}_i+\sum_{k=0}^{N} K_0\left({y_{k+1/2}\over x_i}
	\right)\left[(y_{k+1}\epsilon_{k+1}-y_k \epsilon_k)
	+{\epsilon_{k+1}^2-\epsilon_k^2 \over 2}\right].
\label{e41b}
\end{equation}
The terms ${\cal C}_i$ and ${\cal D}_i$ are needed
to correct for the part of the integrations from $y_{N+1}=x_{N+1}$
to $\infty$, which we perform under the assumption that
$\epsilon(y)=\epsilon_{N+1} (y/y_{N+1})^{-2}$ in these regions:
\begin{equation}
{\cal C}_i = -2\epsilon_{N+1}
h\sum_{k=N+1}^\infty K_0\left({y_{k+1/2}\over x_i}\right)
{y_{N+1}^2 \over y_{k+1/2}^3} ,
\label{e42a}
\end{equation}
\goodbreak
\begin{equation}
{\cal D}_i = -h\sum_{k=N+1}^\infty K_0\left({y_{k+1/2}\over x_i}
\right) \left( {\epsilon_{N+1} y_{N+1}^2 \over y_{k+1/2}^2} +
{2\epsilon_{N+1}^2 y_{N+1}^4 \over y_{k+1/2}^5} \right) . 
\label{e42b}
\end{equation}
The summations (\ref{e42a}) and (\ref{e42b}) can 
be performed in advance for all enclosed
grid points $i=1,\dots, N$ without introducing significant errors 
if we choose a sufficiently large number of extra terms beyond $N+1$.

The important part of the procedure comes from placing the source point 
$y_{k+1/2}=(k+1/2)h$ mid-way between any two field points $x_i$ so 
that $K_0(y_{k+1/2}/x_i)$ is never evaluated at the singularity of 
$K_0(y/x)$ at $y=x$. For this reason, no extra ``softening'' of the 
kernel is needed. One can in principle choose a source point that is 
not exactly mid-way between two field points. In such a case, the 
replacement of the (principal-value) integrals, $\cal I$ and $\cal J$, by the sums (\ref{e41a}) 
and (\ref{e41b}) would be only first-order accurate, and they
would not take proper advantage of the cancellations
resulting from the change of sign of $K(y/x)$ as the integration over
$y$ occurs across the field point $x$. To equation (\ref{e40a}), we wish to add the two BCs:
\begin{equation}
\epsilon_0 = \surd 2, \qquad \epsilon_{N+1} = 
\epsilon_N \left( 1+{h\over N}\right)^{-2}. 
\label{e43}
\end{equation}
The trick to obtaining a good numerical solution is now to compute
${\cal R}_i$ from equation (\ref{e40b}) as a known quantity from previous 
iterates for $\epsilon_k$.  New iterates are obtained then by solving the
{\it linear} set of simultaneous equations (\ref{e40a}) and (\ref{e43}) 
for the $N+2$ variables: $\epsilon_i$ for $i=0, 1, \dots, N, N+1$. The 
number of interior grid points $N$ needed for an accurate solution 
depends on the value of the outer radius $x_{N+1}$ chosen. We find 
that as long as the values of the grid size $h$ are less than about 
0.1, the converged solutions are practically indistinguishable for one 
another. 

\subsection
{Tridiagonal Matrix Inversion, Relaxation Technique, and Initial Iterate}

Equation (\ref{e40a}) and the boundary conditions (\ref{e43}) can be cast 
into a tridiagonal matrix equation.  We solved this equation with the
subroutine TRIDAG from ``Numerical Recipes'' by Press et al. 
(1986). Because we do not use a full linearization Henyey-technique,
we add a relaxation step to ensure convergence.
If $O_k$ represents the old iterate for
$\epsilon_k$ with which we calculate ${\cal R}_i$, and $N_k$ 
represents the
new solution for $\epsilon_k$ that we get
by solving the matrix equation, we define the new iterate as
\begin{equation}
\epsilon_k = \alpha N_k + (1-\alpha) O_k \qquad {\rm for} \qquad
k = 0, 1, \dots, N, N+1 
\label{e44}
\end{equation}
where $\alpha$ is a relaxation parameter.  Notice that
equation (\ref{e44}) maintains $\epsilon_0$ at the 
inner-boundary value $\surd 2$, 
but $\epsilon_{N+1}$ will vary as the interior variables are updated.
A choice of $\alpha$ between 0 and 1 corresponds to under-relaxation;
a choice greater than 1 corresponds to over-relaxation; and
a choice less than 0 corresponds to liking old iterates better 
than new ones.  For the converged numerical solution to be shown, 
we find that a small value of $\alpha=0.001$ is needed to avoid
instabilities in the iteration. 

The only remaining chore is to make a reasonable initial guess for
$\epsilon(x)$.  We choose the function:
\begin{equation}
\epsilon(x) = {2\over \surd 2 + x+ 2x^2} ,
\label{e45}
\end{equation}
which has the desirable properties of being everywhere positive,
with $\epsilon(0)=\surd 2$,
$\epsilon^\prime (0) = -1$, and $\epsilon(x)\rightarrow
1/x^2$ at large $x$. (The coefficient 1 is chosen with some knowledge
of the numerical solution).  

By systematically increasing the value of the last computational 
point $x_{N+1}$, we may check that the power-law decay, $\epsilon(x)
= A_2 x^{-2}$, is indeed the correct asymptotic behavior for the 
residual enclosed mass $\epsilon(x)$.  

\subsection{Numerical Solution}

We carried out a numerical determination of the eigenfunction 
$\epsilon(x)$ corresponding to the collapsed final state with 
$m_*=\surd{2}$ in a region between the origin and a large outer
radius of $x=100$. The result is plotted in the top two panels of
Figure 2.  Recall that 
$\epsilon$ denotes the difference in cylindrically enclosed mass 
between the final and initial state. The first panel of Fig.~2 from the top 
shows that the final mass distribution differs
substantially from that of the initial state only in the region 
$x\lsim 2\surd 2$
(i.e., $\varpi\lsim 2^{3/2} r_0$).  The part $x\le \surd 2$ is strongly
affected because this originally supercritical to moderately subcritical
part of the initial cloud fell into the origin to make the star.
The resulting hole has to be filled in by a comparable amount of
matter, which extends the region of influence to $x\sim 2\surd 2$,
given that enclosed
mass scales with radius in the original configuration.  Beyond $x\gsim
2\surd 2$, the background field
lines are too rigid to allow much horizontal motion.
The second panel of Figure 2 shows
that in the strongly subcritical envelope, the 
excess enclosed mass $\epsilon(x)$ approaches zero,
roughly as a power-law $A_2/x^2$ for large $x$ ($\varpi \gg r_0$).
The coefficient $A_2$ has a numerically-determined value of $1.1$. 

The third panel of Figure 2 shows the total enclosed mass $m(x)$
plotted against $x$.  Notice that $m(x)$ has a flat basin
for a range of $x \lsim 1$ 
outside the origin before $m(x)$ smoothly joins the linear relationship
$m(x)=x$ characteristic of the unperturbed state.
The flat basin indicates that
the region where the repulsive force associated with the
split monopole and the attractive gravitation of the central mass point are
in approximate balance extends over a healthy range of $x$.  Thus, equation
(\ref{e28a}) is more than a requirement valid only
just outside the stellar surface
(roughly $x\sim 10^{-5}$ for practical applications).  This is fortunate
as we would otherwise not be able to resolve the criterion (\ref{e28a})
via numerical simulation.  The 
precarious act of extended levitation is accomplished by
stretching out the material with a mass-to-flux ratio
nearly given by $\lambda \approx 1/\surd 2$.
For $x \gsim 1$, the gravity of the point mass has
dropped sufficiently (as $2^{1/2} x^{-2}$) 
that a {\it negative} magnetic suspension 
must help the gravity of the point mass oppose the outward
expansive forces of magnetic levitation.
In the outer envelope, $x \gg 1$, the balance is almost entirely
between positive levitation and negative suspension.  This is
shown quantitatively in the fourth panel of Figure 2,
where the combination ${\cal J}(x)+1$, proportional to the combination
of levitation and suspension (i.e., the total magnetic tension force)
rapidly approaches zero as the field lines
become straight and uniform at large $x$.  
The other forces of gas pressure and self-gravity
in the envelope as $x\rightarrow \infty$
are basically left to find their own balance,
as was true in the initial state.

To see these balances in another way,
we plot in Figure 3 the distributions of the mass column-density $\Sigma$
in units of $B_0/2\pi G^{1/2}$ and the vertical component of the magnetic field
$B_z$ in units of $B_0$ at dimensionless radius $x=\varpi/r_0$. 
The former is computed from equation (\ref{e34}) 
and the latter from equation (\ref{e12}).
For $x\lsim 2\surd 2$, $\Sigma$ lies significantly
below its initial value (dashed curve depicting $1/x$), because --
after all -- half of the material that used to be here has fallen
into the central star, and the other half must be redistributed over
the ``hole'' that's been left behind.
The redistributed mass shows a much shallower rise toward the origin
than we might have naively expected for a gravitationally
collapsed region, with an
inner drop inside $x\approx 1$ that seems totally mysterious
at first sight.  For $B_0$ = 30 microgauss, the shallow rise
and fall toward the center produces dimensional column densities
$\lsim 9\times 10^{-3}$ g cm$^{-2}$
that cosmic rays, or even stellar X-rays, would have no problems
penetrating to produce sufficient ionization
for good magnetic coupling throughout the region of interest
(cf. Nishi et al. 1991, Glassgold et al. 2000).  Indeed,
the column density approaches the level where the general ultraviolet radiation
field of the interstellar medium provides significant ionization
(McKee 1989).

The gentle mound of matter piled up toward the center
arises because material with
subcriticality $\lambda \approx 1/\surd 2$ is being
pulled by a star of dimensionless mass $m_*=\surd 2$ and pushed by a
split monopole of comparable strength.   This stretching
produces the flat basin of $m(x)$ mentioned
in the previous paragraph. The pull dominates in the interior and
the push dominates
in the exterior, which explains why the mound has a central depression.
In turn, this stretching of the matter distribution pulls and pushes
the footpoints of the magnetic field threading through the sheet
in such a way
that the magnetic field $B_z$ has everywhere
(except for the central star) magnitude {\it less than} 
the background value $B_0$
(horizontal dashed line in dimensionless units).
This produces a contradiction with maser measurements
of magnetic field strengths of collapsed star-formation
regions (Fish \& Reid 2003) that we return to resolve in \S 4.
In any case, the negative value of
$B_z-B_0$ is what produces the negative suspension force of the cloud envelope
needed to counteract the positive levitation of the split monopole
(see eq.~[\ref{e9a}]).

In summary, the numerical solution shows that
the split monopole of the star more than just holds up the pseudodisk
in the immediate vicinity of the origin.  It also
pushes up against the background magnetic field off the midplane that 
threads through the pseudodisk (strongly perturbed part of the gas 
distribution
outside the star) and cloud envelope (weakly perturbed parts of the initial
state) and is distorted by such fields.  In back-reaction, the 
background magnetic field
is distorted in such a manner that suspension forces
help to hold the gas envelope, 
{\it not out}, but {\it in} (see Appendix B).

\section{Time-Dependent Evolution of Initial State to Final State}

We now discuss the time-dependent, axisymmetric, numerical simulations that we
performed for a singular isothermal sphere (SIS) threaded initially
for a uniform magnetic field in the $z$ direction.  The calculations were
done with a modified version of Zeus2D (Stone \& Norman  1992a,b).  The
modifications are described in Allen et al. (2003a).

At time $t=0$, we initiate the inside-out collapse by adding
a small seed mass in the central cell of the calculation.
The resulting (non-dimensional) mass-infall rate
$\dot m \equiv G{\dot M}/a^3$ is shown
in Figure \ref{fig:Mdot} as a function of dimensionless time
$\tau \equiv at/r_0$.  The values of $\dot m$ at small $\tau$
are close to the unmagnetized SIS value of 0.975.
For $\tau \gsim 1$, as the wave of
expanding infall engulfs ever more subcritical envelope material,
the infall rate declines steadily from 0.975.
By $\tau = 5.4$, the dimensionless mass $m_*\equiv M_*/M_0$
in the central cell has accumulated 56\%
of its final value of $\surd 2$ anticipated from the
2-D axisymmetric analysis of \S 3.  At this
time, when the flow is trying to establish a delicate
balance between magnetic levitation by
the split monopole and the gravitation of the point mass
at the center, numerical diffusion associated with
the artificial viscosity in the code introduced to mediate
accretion shocks (and other truncation errors) sets off growing oscillations that
resemble those which plagued the late-time simulations of
Allen et al. (2003a,b).  The manifestation of these oscillations
in the instantaneous mass accretion rate for $\tau > 5.4$
can be seen in Figure \ref{fig:Mdot}.  An extrapolation of the
time-averaged behavior of $\dot m$ suggests that
the final accumulated mass $m_*$ would be consistent with the
theoretical value of $\surd 2$.  Indeed, much of the infall into
the center originates at late times from material high off
the midplane flowing down field lines that have been pulled
into the central cell by the gravitational collapse, and
the total amount of such material is finite.

Figure \ref{fig:flow} depicts the flow
in the meridional plane at $\tau = 0.1, 1.0, 2.5,$ and $5.4$
before serious numerical oscillations commence.
The solid curves of each snapshot show the configuration
of magnetic field lines; arrows depict the local flow direction;
and isodensity contours are given by heavy dashed curves, with
the densest portions highlighted by shading.
For each frame, a split monopole emerges from the central cell,
which pushes out the field lines threading an
inflowing pseudodisk before the split monopole
fields straighten to attach to the imposed background fields.
The fields threading the midplane of the configuration, in turn,
progressively push on their neighbors radially farther out to give
the magnetic tension forces that suspend the outer envelope against
the part of the
gravity of the central mass point and the pseudodisk
that is not resisted by the gas-pressure gradient.
The field strengths throughout the regions of interest (apart from
those that go through the central zone) are everywhere equal to or
below background values.
Except for the vertical flare of the flattened
pseudodisk into a toroid-like
configuration at large $x\equiv \varpi/r_0$, the pictures in Figure
\ref{fig:flow} for late times
are much as we would have anticipated on the basis
of the axisymmetric 2-D final-state analysis of this paper.

Figure \ref{fig:sigmaB3} lends additional weight to the
correspondence between the 2-D and 3-D axisymmetric calculations.  Plotted
here against the dimensionless radius $x\equiv \varpi/r_0$
at the midplane of the calculation are the projected mass density
$\Sigma$ and vertical magnetic field $B_z$ (the only nonzero
component at $z=0$) at a time $\tau = 8.5$, when the accumulated
central mass is within 66\% of its expected final value of $\surd 2$ times
$M_0$.  The units for $\Sigma$ and $B_z$ are
$B_0/2\pi G^{1/2}$ and $B_0$, respectively, the same as used
to make the graphs in Figure \ref{fig:sigmaB} for the final state
in the 2-D axisymmetric calculations.  Allowing for the distortion of the central
regions because of the numerical oscillations, we see that the 3-D axisymmetric theory
at late times holds no surprises that could not be anticipated by the 2-D axisymmetric
analysis.

\section{Discussion and Summary}

\subsection{A Powerful Conjecture}
\label{Conjecture}

Equations (\ref{e28b})--(\ref{e31}) give the most salient features
of the formal solution for the ``initial'' and ``final'' states of
the posed collapse problem in axisymmetric 2-D
if we insist on strict field freezing.  
It is interesting that the central masses that
result from the dimensional formula,
\begin{equation}
M_* = \surd 2 M_0 = \pi^2 \surd 2 {a^4\over G^{3/2} B_0},
\label{finalmass}
\end{equation}
can be made to correspond to stellar masses only if we choose
values of $a=(kT/\mu)^{1/2}$ and $B_0$ that correspond to interstellar
values, yet the physical determination of the mass eigenvalue
occurs close to the star, where the local temperature $T$ and $B_z$
in the realistic situation are likely to differ considerably
from their interstellar counterparts.  How sensitive, therefore,
are our numerical conclusions to the exact details of the
model adopted to do the actual calculations, in particular, to the assumptions
of isothermality (same temperature in initial and final state) and
field freezing (same $dM/d\Phi$) in initial and final states?

Apart from numerical factors of order unity,
we believe that our conclusions are generic.
The following is a ``proof'' that does not quite live up to the
standards of rigorous mathematics, but is powerful by
astronomical standards.  If there is to be a final equilibrium 
involving a highly flattened, nonrotating,
configuration of matter immediately outside
the star, the final state of equilibrium still has to satisfy equation 
(\ref{e11}).  For realistic temperatures in the pseudodisk (i.e.,
not approaching values comparable to that at the center of the star), 
the pressure gradient term in equation (\ref{e11})
is negligible in comparison the stellar gravity term $-M_*$ 
throughout the inner pseudodisk.
Similarly, as long as the mass contained in the
pseudodisk from the stellar surface
out to radius $\varpi$
is less than the stellar mass $M_*$ multiplied by $\lambda(\varpi)$
times $\bar \lambda$ where $\bar \lambda$ is some appropriate average
value in the pseudodisk, 
the contribution from the integral is negligible in comparison
with the stellar gravity term.  (This is always true for
$\varpi$ near the inner
edge $R_*^+$.)  Thus, the only way to satisfy equation (\ref{e11})
is to have the split-monopole levitate the
inner pseudodisk against the stellar gravity,
i.e., the mass-to-flux ratio in the inner pseudodisk must have 
an inverse relationship to the stellar mass-to-flux ratio:
\begin{equation}
\lambda_*\lambda(R_*^+) = 1 . 
\label{eLL}
\end{equation} 
This elegant equation is the generalization of our eigenvalue 
determination for the final-state problem.

Suppose now that the initial cloud that gave rise to the collapsed
final state has a flux-to-mass distribution given by
\begin{equation}
{d\Phi\over dM} = 2\pi G^{1/2} f^\prime \left({M\over M_0}\right),
\label{dPhidM}
\end{equation}
where $M_0$ is no longer necessarily given by
equation (\ref{e-4}) but is equal,
say, to the observed value where the cloud turns from being supercritical 
(core) to subcritical (envelope).  We suppose therefore that
the function $f^\prime(m)$ in equation (\ref{dPhidM})
is a monotonically increasing
function of its argument with the properties $f(0)=0$ and $f^\prime(1)=1$.
Assuming strict field freezing in the transition from the initial
to final state, and
carrying out an integration of equation (\ref{dPhidM}),
we find that the flux to mass ratio of the final star is
given by
\begin{equation}
{\Phi_*\over M_*} = 2\pi G^{1/2} f\left({M_*\over M_0}\right){M_0\over M_*}.
\label{PstarMstar}
\end{equation}
Denoting $m_* \equiv M_*/M_0$, and identifying $\lambda(R_*^+)=
[f^\prime (m_*)]^{-1}$ and $\lambda_* = m_*/f(m_*)$,
equation (\ref{eLL}) obtains the general solution for the multiple
$m_*$ of $M_0$ that will fall into the star as the positive root of
the equation:
\begin{equation}
f(m_*)f^\prime(m_*)-m_* = 0.
\label{generaleigen}
\end{equation}
For the special case, $f(m)=m^2/2$, we recover the eigenvalue solution $m_*=\surd 2$ of \S 2.

Equation (\ref{generaleigen}) is a remarkable result.  It claims the 
following.  Start with a magnetized cloud of arbitrary shape and size
(much bigger than a star)
that has a dimensionless flux-to-mass distribution $f(m)$, where
$m$ is the mass $M$ in units of the core mass $M_0$ (the supercritical
part of the cloud) within magnetic flux $f$, which is $\Phi$
in units of $2\pi G^{1/2}M_0$.\footnote{To construct the function $f(m)$
in the presence of turbulence and lack of any exact symmetries, start
with the dimensional flux-to-mass ratio $d\Phi/dM$ computed by projecting the mass
along each (wiggly) flux tube onto some arbitrary
mid-plane at some instant $t=0$.  Define the central flux tube $\Phi=0$ to be the one
that contains the minimum value of $d\Phi/dM$.  Plot as a contour diagram
the values of $d\Phi/dM$ surrounding this central flux tube.  The function
$\Phi(M)$ is obtained by the two-dimensional integration of this contour diagram
from $\Phi=0$ to various isocontour levels of $d\Phi/dM$ that contain flux
$\Phi$ and mass $M$.  The core mass $M_0$ is defined by the isocontour level
where $f'(m) \equiv 2\pi G^{1/2} d\Phi/dM = 1$, with $m\equiv M/M_0$.  With
$f'(m)$ known, a simple integration, with $f(0)=0$, yields $f(m)$.  
Notice that there is no need to define a length scale $r_0$ in such a formalism,
nor do we need to introduce the concepts of a typical isothermal sound speed $a$
or background field strength $B_0$.  Finally, there are no restrictions
on the nature of the initial velocity field of the fluid, laminar or turbulent.
We have assumed that there is only one local minimum to $d\Phi/dM$, i.e., that
all isocontours close around a single point.  The case when a 
common (subcritical)
envelope surrounds many local minima of $d\Phi/dM$ remains to be explored 
(see \S\S 4.3 \&
4.4).}  Then, without doing any calculations
other than finding the root to equation (\ref{generaleigen}), we are able to
predict accurately the final mass of a star from the ideal MHD collapse of this cloud.

The initial conditions could contain an
arbitrary amount of initial rotation, whose elimination
by magnetic braking when field freezing holds -- see 
Allen et al. (2003b) -- will
ensure the assumption of a single rotating star as the final state.
The initial state also needs not be a configuration of equilibrium or near-equilibrium.
It could contain sizable turbulent fluctuations in the density
and velocity fields -- particularly in the magnetically subcritical envelope
--  these will dissipate without having any effect on the final state.
Sufficiently large turbulent fluctuations in the supercritical core
(which are not seen in the cores that give rise to low-mass star formation)
could give rise to temporary fragments, but in the presence of trapped
stellar or pseudodisk flux which will stir the surrounding cloud envelope, the 
fragments are likely
to lose all their orbital angular momentum and merge into a 
single final object.\footnote{For very weakly magnetized initial states,
or special geometric configurations, the time scale for the Alfv\'en radiation
of the system angular momentum to the envelope of the cloud may exceed
relevant YSO time scales.}  Finally,
the thermal history can be arbitrarily complex.  This will
also have no effect on the final state, provided that radiation pressure
on dust grains does not come into play (true for low-mass stars)
and the temperature in the inner pseudodisk does
not approach virial values.

\subsection{Speculations}

For realistic astronomical application, the weakest link in the above argument is,
of course, the assumption
of field freezing.  Indeed, nearly fifty years ago, Mestel \& Spitzer (1956)
recognized that field freezing would predict stars at birth with surface
magnetic fields of many megaguass, contrary to the observational evidence.
They also pointed out (Mestel 1965, 1985; Spitzer 1968) that conservation of angular momentum in the gravitational collapse of stars from interstellar clouds would lead to absurd
rotational velocities of the final objects, again contrary to the observational evidence.
They proposed that a solution be found to link these two problems via magnetic braking
and field slippage.  We now propose to link the two classical problems of
magnetic flux loss and angular momentum redistribution to the
third great problem of star formation: What processes define the
masses of forming stars?  We even believe that the breakdown
of field freezing plays a fundamental role in the fourth great problem of star formation:
What determines whether gravitational collapse produces
binary (or multiple) stars versus single stars surrounded by planetary
systems?  (See, e.g., the discussion of Galli et al. 2000 and 
the simulations of Nakamura \& Li 2003.)

The formal answer given by equation (\ref{generaleigen}) to what determines
stellar masses is magnetic 
levitation.  But levitation by a stellar split monopole leads to
numerically unacceptable surface magnetic fields for the
central star, as well as to the wrong prediction that
magnetic fields should be {\it weaker} in the regions adjacent to
newly formed stars than the surrounding molecular cloud (see Figs. 
\ref{fig:sigmaB} and \ref{fig:sigmaB3} as well as Fish \& Reid 2003).
Is it possible that magnetic reconnection
of the type contemplated by Galli \& Shu (1993b, see also 
Mestel \& Strittmatter 1967)
completely eliminates the stellar monopole 
(except for some small seed fields needed to start stellar dynamo action)?
After all, the strong sheet current corresponding to the reversal of 
field lines across the midplane of a split monopole will dissipate in
the presence of finite resistivity.  The dissipation of the
sheet current corresponds to the annihilation of field lines
just above and below the midplane. The magnetic pressure of remaining
field lines presses toward the midplane to replace the lines that
have been annihilated.  They are annihilated in turn.  The result is an
ever weaker split monopole, which continues to lose strength
by sheet current dissipation
until the sheet current and the split monopole are both gone.  (In
practice, this dissipation would occur throughout the process of central
mass and flux accumulation and not just in the final state.)

A vanishingly small split monopole, $\lambda_*\rightarrow \infty$
in equation (\ref{eLL}), requires $\lambda(R_*^+) \rightarrow 0$
if magnetic levitation is still to do its work.  
The only material so heavily magnetized outside the star
(if magnetic flux were preserved even approximately
in the rest of the system where the fields are not so severely
pinched) comes from far out in the original 
model cloud. In other words, with no split monopole
at the center to help it, magnetic suspension of
the cloud envelope alone cannot withstand the very large gravity 
just above the surface of the star, and the infall would continue until the 
star has accumulated the entire molecular cloud (or stellar evolution 
intervenes).  

But this possibility also contradicts the observational evidence.
Therefore, imagine next that there is no electrical resistivity, but
there is ambipolar diffusion.  Unlike electrical resistivity,
ambipolar diffusion cannot destroy magnetic flux, but it can redistribute it.
Can ambipolar diffusion (or C shocks, see Li \& McKee 1996)
transfer the flux from material before it became incorporated into the star
and redistribute it over a much wider area of a static pseudodisk, so that the
resultant fields are not such an embarrassment relative to observations?  
On a larger scale the levitation provided by such fields could serve
to prevent the infall of the cloud envelope.  The difficulty of such a scheme
is that there is only suspension to keep the static pseudodisk from falling
into the star, and we proved in this paper
that suspension without levitation cannot do
such a job.

Keplerian rotation provides an excellent barrier to infall,
so can the conversion of magnetized pseudodisk into a magnetized
Keplerian disk provide the needed buttress to prevent the inward advance
of the cloud envelope?  The question is especially cogent now that
Allen et al. (2003b) have demonstrated that centrifugally supported disks
cannot arise in realistically magnetized models of rotating gravitational
collapse unless the assumption of field freezing is abandoned in the
pseudodisks and disks of star formation (see, also, Krasnopolsky \& 
K\"onigl 2002 for examples of formation of centrifugally supported 
disks in cases of low, prescribed, efficiency of magnetic braking in
the presence of ambipolar diffusion).   
But strongly magnetized, rapidly rotating
disks will develop magnetocentrifugally driven disk winds
(Blandford \& Payne 1982, K\"onigl \& Pudritz 2000), which will provide
{\it dynamic rather than static} levitation.  Thus, ultimately, the
realistic solution to the problem of obtaining stellar masses 
from the gravitational collapse of large interstellar clouds
will end up invoking wind-limited mass infall (Shu \& Terebey 1984).
Moreover, numerical simulations of Keplerian disks threaded by such
a network of open field lines show them to quickly fall into the
innermost regions of the disk (Gammie \& Balbus 1994, Miller \& Stone 1997), 
concentrating all the trapped flux into a narrow
annulus that makes the subsequent outflow indistinguishable
from an X-wind.

X-winds will also arise in the alternative extreme scenario of
all electrical resistivity and no ambipolar diffusion, if Keplerian disks
appear outside accreting protostars (Shu et al. 1988, 1994, 2000).  In this case,
although the stellar split monopole will eventually disappear
because of field annihilation (eliminating the magnetic braking
by long lever arms that prevents the formation of Keplerian disks
in the simulations of Allen, Li, \& Shu 2003), the operation of a stellar dynamo
can replace such fields with those that have a dipole or higher multipole
structure at the stellar surface (Mohanty \& Shu 2003).  The interaction of these dynamo-driven
fields with the surrounding accretion disk, and the subsequent opening of
the stellar field lines (or the simple pressing of the remnant 
interstellar field lines against the magnetopause) can then
create an X-wind.  The attractiveness of either proposal relative to
that of static levitation is one of efficiency.  Newly formed stars
do not need 10 megaguass fields to halt the inflow of gravitational collapse.
Surface fields of a few kilogauss, when combined with the rapid
rotation from an adjoining accretion disk, suffice to provide
dynamic levitation (a physical blowing away) of cloud envelopes.

In the real situation, electrical resistivity and ambipolar diffusion 
are probably equally important in the disks and pseudodisks around
young stars.  Since we have argued for X-winds
as a likely outcome for the two extreme assumptions, we believe
it is also the likely outcome in all intermediate cases.

\subsection{From Isolated Star Formation to Distributed Star Formation}

Imagine a large magnetized cloud which is overall subcritical,
and therefore probably fairly flattened, but which contains 
many dense pockets (cloud cores) that are supercritical.
(We are supposing that the discussion of
\S \ref{Conjecture} can be generalized to more than a single core.)
Each of the cores can undergo inside-out collapse.
If they later generate a magnetocentrifugally driven wind, 
they can each dynamically levitate their neighboring subcritical envelopes
and prevent star formation from being anywhere near 100\% efficient.
A general inefficiency for forming stars is consistent with the observational evidence,
except possibly for situations involving the bound-cluster mode of star formation.
The supersonic turbulence and other complex factors in the subcritical
common envelope play no
dynamical role once the cores go into collapse
and generate their winds, but the winds
may play a crucial role in sustaining the observed turbulence
and overall filamentary morphology of the cloud (Allen \& Shu 2000).
In turn, as discussed below, the turbulence in the general envelope
may ultimately help set the mass distribution of molecular cloud cores
{\it before} the onset of gravitational collapse, and therefore the
initial-mass-function of forming stars.

How did the cloud get so many supercritical cores in a general environment
that is subcritical?  One answer proposed by Shu et al. 
(1987, see also Lizano \& Shu 1989) is ambipolar diffusion acting as a background process 
in all molecular clouds.  Their name for this process was ``the isolated
mode of star formation,'' but in the present generalization, we
refer to it as ``the distributed mode of star formation.''

If ambipolar diffusion is the mechanism that condenses many cores
from large cloud envelopes, it represents a kind of {\it
magnetic fragmentation}.  Whether this fragmentation occurs slowly or
not relative to dynamical timescales may depend on the contractive and diffusive
enhancements one ascribes to turbulence dissipation and fluctuations 
(Myers \& Lazarian 1998, Myers 1999, Adams \& Fatuzzo 2002, Zweibel
2002).  In any case, the first supercritical core to appear from
the gravitational contraction of any dense pocket of
gas and dust has, by definition, a vanishingly small mass $M_0$,
since the transition from subcritical to supercritical is made gradually.
Unless dynamically compressed,
such a core is not immediately unstable to inside-out gravitational 
collapse; it has to evolve by further ambipolar diffusion
(and perhaps turbulent decay)
to reach a state of sufficient central concentration.
In the numerical simulations performed to date of this process,
the exact state of runaway collapse
depends on the outer boundary conditions
that are imposed
(see, e.g., Nakano 1979; Lizano \& Shu 1989; 
Tomisaka, Ikeuchi, \& Nakamura 1990;
Basu \& Mouschovias 1994; Li 1998).  When those boundary conditions
are applied at infinity, and the evolution is approximated as
quasi-static, our group has argued for pivotal states that resemble
singular isothermal toroids (Li \& Shu 1996, Allen et al. 2003a),
which are close relatives
in their density distributions (but not flux distributions)
to the magnetized SISs studied in this paper.

The details for cloud core structure are not important for
the present application.  Recently, Motte, Andr\'e \& Neri (1998),
Testi \& Sargent (1998), and Motte et al. (2001) have found that the mass
distributions of cloud cores as mapped by dust emission
in turbulent molecular clouds resemble
the empirical initial mass function (IMF) of newly formed stars
(Salpeter 1955, Scalo 1986).  The weighting of cloud cores toward small
(typically solarlike) masses instead of toward large masses
(typically thousands of $M_\odot$)
as in cloud clumps (sometimes called {\it dense} cores) is particularly
striking.\footnote{We reserve the name ``cloud core" for the
supercritical entities that will form individual stars or binaries.
Observers define such cores differently
than we do, based on observable contrast over the background rather than
on the criticality of the mass-to-flux ratio.
For realistic core-envelope models, the two
definitions are related, since being supercritical allows the self-gravity
of a region to produce large column density contrasts compared with
their surroundings, which are probably subcritical or only weakly
supercritical.}
Lada \& Lada (2003; see also Lefloch et al. 1998, and Sandell \& Knee 2001)
make the interesting suggestion that outflows may
transform the mass-spectrum of cloud clumps, which resembles
the mass distribution of embedded clusters, to the mass-spectrum of cloud cores,
which resembles the mass distribution of stars.
In what follows, we suggest a mechanistic process for performing
such a task in a statistically invariant way, both in the distributed and
the clustered modes of star formation, if outflows ultimately provide a major
source of the turbulence present in star-forming clouds.

\subsection{Initial Mass Function}

As an example of how the processes described in this paper can determine the
stellar IMF (see also Silk 1995 and Adams \& Fatuzzo 1996), we
consider the contribution lent by turbulent velocities $v$ to the support
of molecular clouds.  At a heuristic level, we imagine that $v$ can
be added in quadrature to the isothermal sound speed $a$ in formulae that
involve the gravitational constant $G$ (e.g., Stahler, Shu, \& Taam 1980;
Shu et al. 1987; McKee 2003; Pudritz 2003).  For even greater simplicity, we use $v$ 
below to mean the turbulent velocity when the turbulence is supersonic,
and we substitute $a$ for $v$ when that turbulence is subsonic.

From IRAM and CSO observations of molecular clouds, Falgarone \& Phillips
(1996) find that CO line-profiles have high-velocity wings indicative of
supersonic turbulence.  This interpretation is supported by the study
of Falgarone (2002) who argues that the structure of molecular clouds
is characterized by a unique power-law in mass,
except possibly for a lower scale defined by cloud cores
(Blitz \& Williams 1999).  We adopt such a description and
assume that the mass of magnetized molecular-cloud material with
turbulent velocity between
$v$ and $v+dv$ in 1-D is given by the power-law distribution,
\begin{equation}
{\cal M}(v)\, dv \propto v^{-s} \, dv,
\label{turbvspectrum}
\end{equation}
where $s$ is a positive number.  To avoid divergence of the 
integrated mass at very small $v$, we cut
off the distribution (\ref{turbvspectrum}) when
$v$ becomes smaller than the thermal line-width $a$,
noting that the important applications occur at $v\lsim 1$ km s$^{-1}$.

To ensure that there is no confusion over what we mean by ${\cal M}(v)\, dv$,
we emphasize that we have in mind a statistical sampling of a whole giant molecular
cloud (GMC), or even of an ensemble of GMCs in an entire galaxy.  As a practical
way to measure ${\cal M}(v)\, dv$, at least at relatively large $v$'s (a few
km s$^{-1}$), imagine getting the integrated spectrum of a GMC in the line wings of
some optically thin species, after we remove the effects of a systematic velocity
field, and determining the mass with random velocity between $v$ and
$v+dv$ along the line of sight.  Then our derivation below holds
assuming that the same functional form ${\cal M}(v)$ extrapolates
to $v \lsim 1$ km s$^{-1}$ and to the material in molecular cloud cores.

With this understanding, we further suppose that the core mass
$M_0$ can be approximated as (see eq.~[\ref{e-4}]):
\begin{equation}
M_0 = m_0{v^4\over G^{3/2} B_0},
\label{m0M}
\end{equation}
where the coefficient $m_0$ has a probabilistic distribution centered about
$\pi^2$ with a variance of a factor of a few (see \S 4.6).
Values of $m_0$ smaller than the typical $\pi^2$ arise because
turbulence, if compressive rather than expansive, can act to decrease as well
as to increase the stability of some fiducial statistical equilibrium.
Suppose now that the final stellar mass is some
fixed fraction of $M_0$ (say, 1/3; see \S 4.6): 
\begin{equation}
M_* \sim {1\over 3} M_0 \sim {m_0 v^4\over 3G^{3/2}B_0}.
\label{finalmscaling}
\end{equation}
With a stellar mass-accumulation
rate behaving as (2/3 times the infall rate) 
$\dot M_* \sim 2v^3/3G$ (cf. Shu et al. 1987),
we get a characteristic formation time,
\begin{equation}
t_{\rm sf} \equiv {M_*\over \dot M_*} \sim {m_0 v\over 2 G^{1/2}B_0} 
= {m_0^{1/2} r_0\over 2v}, 
\label{tsf}
\end{equation}
where we have adopted equation (\ref{e-3}) with $a$ and $\pi$ replaced,
respectively by $v$ and $\surd m_0$ to estimate the core radius $r_0$.  The infall rate
could exceed $v^3/G$ by a factor of a few if static magnetic fields contribute
to core support before collapse, and if pivotal states contain
nonzero contraction velocities (Allen et al. 2003a, b).
The characteristic time $t_{\rm sf}$ would then
typically measure $\sim 10^5$ yr, rather than a few times this value.

We further suppose that $B_0$ scales with $v$ as
\begin{equation}
B_0 \propto v^p,
\end{equation}
where $p$ is again a positive number.
The mass of stars formed during any fixed interval of time $t_{\rm sf}$
with stellar masses between $M_*$ and $M_*+dM_*$ is now
given by
\begin{equation}
M_*{\cal N}(M_*)\, dM_* =\varepsilon {\cal M}(v)\, dv \propto M_*^{-(s+3-p)/(4-p)} \, dM_*.
\label{IMF}
\end{equation}
In equation (\ref{IMF}), 
$\varepsilon$ is an overall efficiency of turning
molecular clouds into stars and is given by
\begin{equation}
\varepsilon = {F\over 3}, 
\label{SFE}
\end{equation}
where $F$ is the cumulative fraction of molecular cloud mass in the region that has existed as star-forming
cores.  Of the currently observable cores,
a fraction $t_{\rm sf}/(t_{\rm sf}+t_{\rm AD})$ exists as cores with actively accreting
embedded protostars, and a complementary fraction 
$t_{\rm AD}/(t_{\rm sf}+t_{\rm AD})$ exists
as starless cores.  In the latter expressions, $t_{\rm AD}$ is the time required
for ambipolar diffusion and the decay
of turbulence to transform a core with a critical level of magnetization
to one in a pivotal state at the onset of protostar formation.

Except for the normalization factor $\varepsilon$, where we have adopted the observer's definition of
star formation efficiency as the mass of formed stars divided by the mass
of total starting material in the region, equation (\ref{IMF}) corresponds
to Salpeter's (1955) IMF for low- and intermediate-mass stars if
\begin{equation}
{s+3-p\over4-p} = 4/3, \qquad i.e., \; {\rm if} \qquad s={7-p\over 3}.
\end{equation}
In the above, we have taken the liberty of approximating Salpeter's 1.35
by $4/3$.  One attractive combination satisfying the 4/3 rule is $p=1$ and $s=2$.
This case corresponds to equal $t_{\rm sf}$ for stars of all masses (eq. [\ref{tsf}]) 
and Lorentzian line wings (eq. [\ref{turbvspectrum}]).  

From observation (Masson \& Chernin 1992) and theory (Li \& Shu 1996),
it is known that bipolar outflows produce
swept-up mass distributions that satisfy equation (\ref{turbvspectrum}), with
$s$ close to, but perhaps slightly smaller than 2,
except at the highest velocities, where $s$ can be significantly larger
than 2 (Lada \& Fich 1996). It is not known, however,
whether the same distribution applies after the gas has decelerated from velocities
of order 10 km s$^{-1}$ to those of order 1 km s$^{-1}$.
The assumption that $B_0$ scales linearly with $v$ ($p=1$) agrees with the
estimate by Myers \& Fuller (1993) that stars of all masses take $\sim$ (a few times)
$10^5$ yr to form once gravitational collapse starts.  A positive correlation will exist between
$B_0$ and $v$ if regions with larger turbulent velocities require supercritical cores
to reach higher mean densities before they become self-gravitating, resulting in a
greater compression of the mean magnetic field threading the core. 
The Salpeter slope in
equation (\ref{IMF}) does not depend sensitively on the exact choice for $p$ if $s=2$.

In this paper, to match the properties of Taurus cores, we
have adopted a normalization of $B_0 = 30 \; \mu$G
when $v=a=0.2$ km s$^{-1}$.  With this normalization, the minimum stellar mass
to result from equation (\ref{finalmscaling}) is nominally 0.5 $M_\odot$ if we
take $m_0$ to have a typical value of $\pi^2$.  In actual
practice, 0.5 $M_\odot$ is close to the flattening of the observed Galactic 
IMF for
$M_*{\cal N}(M_*)$ (Scalo 1986), but significant numbers of stars form
in the Orion Trapezium region down to and below the hydrogen-burning limit $0.08$ $M_\odot$
(cf. Fig 10 of Lada \& Lada 2003).  In the current physical
picture, stars can form with masses smaller than 0.5 $M_\odot$,
because there is a distribution in the value of
$m_0 a^4/B_0$ for different regions.  In this context, it may be meaningful that
Taurus appears to be deficient in brown dwarfs in comparison
with the Orion Trapezium (Luhman 2000), consistent with our earlier suggestion
that the effective value of $B_0$ may be statistically lower in the former region.
But it could be that a significant number of brown dwarfs are initially
formed as companions to normal stars and that there are larger numbers of
``free-floating'' brown dwarfs in Orion than in Taurus because of the greater
role of binary disruption in the cluster environment (Kroupa 1995).
It could also be that a limited form of Hoyle's (1955) fragmentation picture
can still apply to turbulent, magnetized, cores if the eventual decoupling of magnetic
fields occur sufficiently rapidly in realistic non-ideal MHD (Galli et al. 2000,
Nakamura \& Li 2003). 

Apart from the different behaviors at low stellar masses,
departures from a $-4/3$ power law in the observed IMF 
appear also at the highest stellar masses.
We believe that an exponent for $M_*{\cal N}(M_*)$
steeper than $-4/3$ can arise at large stellar masses because radiation pressure acting
on dust grains aids YSO winds to reduce $M_*$ as a fraction of the initial core mass $M_0$
(Wolfire \& Cassinelli 1987, Jijina \& Adams 1996).
If this interpretation is correct, the non-power-law features in the stellar IMF
contain clues on how stars help to limit their own masses by blowing away
cloud material that might have otherwise fallen gravitationally into the stars
(cf. Adams \& Fatuzzo 1996).  Figure 7 depicts pictorially
the ideas of this subsection.

\subsection{Embedded Cluster Formation}

Our derivation of the IMF formally seems to
hold only for the distributed mode of starbirth,
where competitive accretion does not
occur.  But apart from the deficit of brown dwarfs, the IMF of YSOs
does not appear to be different in Taurus compared to clusters
or the general field (Kenyon \& Hartmann 1995).
Moreover, the sometimes evoked picture of protostars growing by
Bondi-Hoyle accretion as they move freely in a background of
more-or-less smooth clump gas may be flawed.  After all,
stars do not appear half formed from nowhere; they
probably have to grow by gravitational collapse and infall of small dense cores
that are themselves self-gravitating substructures in the cloud clump.
The gravitational potential associated with the cores forms local minima
that are sharper, although perhaps less deep in absolute terms, than
the large bowl that represents the smoothed gravitational potential
of the clump and associated embedded cluster.  The tidal forces of
the latter will not rip asunder the small cores if their mean densities are
appreciably larger than the mean density of the background clump/cluster gas
(which is the so-called Roche criterion).  

In this regard, it is informative to note that
the mean density inside a {\it sphere} of radius $r_0$ of a uniformly magnetized
SIS is given by
\begin{equation}
\bar \rho_{\rm core} = {2a^2 r_0/G\over 4\pi r_0^3/3} = {3B_0^2\over 2\pi^3 a^2}.
\end{equation}
For $a=0.2$ km s$^{-1}$ and $B_0 = 30 \; \mu$G, $\bar \rho_{\rm core}$
has a numerical value $1.1\times 10^{-19}$ g cm$^{-3}$ $\approx$ 
$1.6\times 10^3$ $M_\odot$ pc$^{-3}$.  This core density is larger than the average
density of any observed clump forming an embedded cluster in the local GMCs
of our Galaxy, except perhaps for the central regions of the very densest clumps
(Lada \& Lada 2003, Table 1). In the central regions of the densest clumps,
individual small cores may merge, yielding the large cores that give rise to
massive stars.  (We presume that such merger processes and large cores are
part of the ``turbulent'' spectrum in eq.~[\ref{turbvspectrum}].)
However, the actual part of the core that forms a star is on average 
9 times denser yet than the value ${\bar \rho}_{\rm core}$, so forming low-mass stars
can survive the neighboring appearance of luminous high-mass stars, although
the remnant envelopes, pseudodisks, and disks not truncated
by the tides of the central cluster/clump, may be photoevaporated away
by the ultraviolet radiation field of the H II regions in which they are embedded
(Johnstone, Hollenbach, \& Bally 1998).   In contrast, the low-mass stars being born
at the centers of the small cores at the peripheries of the clump,
are in no greater danger of wandering away
from those centers than we risk falling off the Earth because it orbits
the Sun with a far deeper gravitational potential.

We do need to worry about the disruption of the cores from hydrodynamic effects
as they orbit inside the clump/cluster.  However, such orbital effects
are probably mitigated by the magnetization of the clump and cores.  This magnetization
{\it dilutes} the effect of gravitation (if both clump and core are supercritical);
and the net forces can actually turn repulsive if the clump (common envelope of
the cores) is subcritical.  Thus, the orbital motions of cores are likely
to be sub-virial with orbital times to cross the clump in excess of the
time $t_{\rm sf}$ required to form individual stars.  The individual cores
may hold together better as entities than envisaged
in the standard picture of a highly turbulent and chaotic clump.

As long as the growing protostar (which feels only gravitational
forces if it completely destroys its accreted interstellar flux by magnetic reconnection)
is trapped by the local potential minimum of its parent core,
the situation will resemble, to zeroth order, the case of isolated star formation.
Sooner or later, such a protostar will
develop an X-wind and begin to blow away its placental core.  The fact
that the common envelope in a cluster environment may be mildly supercritical,
instead of subcritical, may not make too big a difference on our
(admittedly crude) estimate that only 1/3 of the initial supercritical core
will end up falling onto the star (see below).  Indeed, in the present context, the
1/3 figure takes on added significance.  As an extreme, we may think
of a turbulent clump which is supercritical everywhere as {\it completely} packed with
cores, with no surrounding ``common envelope'' (packed like ``kernels'' [of corn on the
cob] in the nomenclature of P. C. Myers).  In that case, the efficiency
of wind-limited star formation for the clump is the same as the average efficiency
for a typical core, which we have taken to be 1/3 (see eq.~[\ref{finalmscaling}]).
The same result, $\varepsilon = 1/3$, can be
obtained from equation (~\ref{SFE}) by setting $F=1$.
It is then interesting to
note that the maximum star-formation efficiency $\varepsilon$ deduced
by Lada \& Lada (2003, Table 2) for embedded star clusters is indeed 33\%. 

To zeroth order, therefore, the clustered mode of star formation under present-day
levels of magnetization may be considered as an extreme form of the
distributed mode of star formation, 
which is itself a generalization of the isolated mode of star formation.
Otherwise, it becomes a total coincidence that the IMFs found
in many different environs of star formation, crowded or dispersed, are
similar, with departures, if any, only at the lower end of the mass spectrum (Lada \& Lada
2003). 

\subsection{Conclusion}

For the mass scale $M_0$ of cloud cores to play its part in
setting stellar masses, cloud envelopes need to be not highly
supercritical.  Indeed, for highly supercritical (e.g., unmagnetized) and turbulent clouds
it is difficult even to define exactly what one means by a ``core'' or ``clump'' except by Roche-lobe
or ``sphere of influence'' arguments.
Moreover, without outflows to halt infall, artificial numerical means to pump in local
turbulence (Klessen 2000) in combination, perhaps, with the ``Jeans swindle'' (Gammie 2003) is needed
to keep the star-formation efficiency of supercritical regions that are overall gravitationally bound
from approaching unity, contrary to the observational evidence.
To understand the overall
low efficiency of star formation, it would be best if the bulk of molecular clouds,
except for their cores, were subcritical or only marginally critical.
Unfortunately, observers have found no regions of molecular clouds that are 
definitely subcritical (Crutcher \& Troland 2000).  This may be because they are biased
to looking at regions where star formation has already taken place,
and such regions are necessarily supercritical.  Allowing for
various projection corrections, Shu et al. (1999) argued that most
Zeeman measurements to date are consistent with the presence of
a near-critical level of magnetization in molecular clouds.
This leaves us optimistic about the fundamental
correctness of the basic program begun by Mestel and Spitzer (1956)
nearly fifty years ago.

Simple numerical estimates suffice to make our case here.  On average,
the envelopes of GMCs possess 4 magnitudes of visual extinction
(e.g., Blitz \& Williams 1999).  With the usual caliberation
that a number-column density of hydrogen $N_{\rm H} = 1.9\times 10^{21}$ cm$^{-2}$
yields 1 mag of visual extinction and that a helium atom accompanies
each 10 hydrogen atoms, the observed extinction through envelopes
corresponds to a mass-column density equal to
\begin{equation}
N_{\rm H} (1.4)m_{\rm H} = 0.018 {\rm g} \; {\rm cm}^{-2}.
\label{GMCenv}
\end{equation}
Zeeman measurements yield the component of the magnetic field along a typical line
of sight through a GMC as $B_{\rm los}$ = 10 $\mu$G (Crutcher \& Troland 1999).
This combination gives an apparent dimensionless ratio of
magnetic field to mass-column density along the line of sight equal to
\begin{equation}
{B_{\rm los}\over 2\pi G^{1/2} N_{\rm H} (1.4 m_{\rm H})} = 0.34
\label{los}
\end{equation}
in the envelopes of GMCs.  For a flat sheet of surface density $\Sigma$
threaded perpendicularly by a magnetic field $B_z$, viewed at an angle $\theta$
with respect to the normal to the sheet, $B_{\rm los}
= B_z\cos \theta$ systematically underestimates the true field
strength $B_z$, and $1.4 m_{\rm H} N_{\rm H} = \Sigma/\cos\theta$
sytematically overestimates the true surface density $\Sigma$.  Since
$\cos^2\theta$ averaged over a hemisphere equals 1/3,
we see that equation (\ref{los}) is close to the expectation value for a sheet
which is exactly critically magnetized:
\begin{equation}
{B_z\over 2\pi G^{1/2}\Sigma} = 1.
\label{envelope}
\end{equation} 

Without bothering to argue about whether better numerical values or better
corrections might make the right-hand side of equation (\ref{envelope})
greater or less than unity, we will merely reiterate
Shu et al.'s (1999) observation that it is probably no coincidence that modern-day
GMCs have near-critical levels of magnetization.  Highly subcritical clouds
are too non-self-gravitating to become molecular clouds; they are present in galaxies
as H I clouds (Heiles 2003).  On the other hand, highly supercritical clouds are too
vulnerable to gravitational collapse; they have long
since disappeared to become the interiors of stars.
McKee's important point that the ultraviolet photons associated with the interstellar
radiation field suffice to keep regions with less than
4 mag of visual extinction too highly ionized to permit  
appreciable ambipolar diffusion then leads naturally to the idea of
magnetically regulated star-formation (McKee 1989).

As a consequence, ambipolar diffusion can act to produce distinctly supercritical
regions only in the cores of molecular clouds.  Many simulations (e.g., Lizano
\& Shu 1989, Basu \& Mouschovias 1994) show that such cores of mass $M_0$, equatorial
radius $r_0$, and mean magnetic field $\bar B$ become susceptible
to gravitational collapse once the dimensionless mass-to-flux ratio acquires a value
of about 2:
\begin{equation}
\bar \lambda \equiv {2\pi G^{1/2} M_0 \over \pi r_0^2 \bar B} \approx 2.
\label{lambda2}
\end{equation}
The dilution of gravity $(1-\bar \lambda^{-2})=3/4$ (Shu \& Li 1997) associated with
this mean level of magnetization is not severe, and
if the same core is supported against gravity largely by turbulent velocity $v$
(as measured in any single dimension), then virial equilibrium requires
\begin{equation}
2\cdot {3\over 2} M_0 v^2 \approx {GM_0^2\over r_0}.
\label{virialequil}
\end{equation}
Solution of equations (\ref{lambda2}) and (\ref{virialequil}) yields
\begin{equation}
M_0 \approx {9v^4\over G^{3/2} \bar B}, \qquad r_0 \approx {3v^2\over G^{1/2} 
\bar B}.
\label{approxMandR}
\end{equation}
When we identify $\bar B$ with $B_0$,
equation (\ref{approxMandR}) for $M_0$ and $r_0$ differ from our expressions
(\ref{e-4}) and (\ref{e-3}) for the same quantities only in the replacement of $\pi$ by 3.

If turbulent cores are barely bound rather than fully virialized -- as
implied by some suggestions that star-forming cores are produced as dynamical
objects rather than as "equilibrium" states -- then the factor
of 2 should be removed from the left-hand side
of equation (\ref{virialequil}).
In fact, since fluctuations can exist on either side of virial equilibrium, we
estimate that $M_0$ given by an expression for star-forming, bound cores such
as equation (\ref{m0M}) can
have a coefficient $m_0$ that might vary by a factor as much as 4 about a
central value of 3 or $\pi$.  This crude estimate is depicted schematically
in Figure 7 showing the typical distribution of $m_0$.

The following reasoning underlies the central idea behind our rough estimate
that the final mass for the formed star is typically one-third
of the core mass, $M_* = M_0/3$.  As long as the infall encompasses
only supercritical regions, the inflow of matter through the envelope
and pseudodisk may be too strong for an incipient X-wind to reverse the infall completely
(see Fig. \ref{fig:Mdot}).  However, when the outwardly expanding wave of infall
begins to encompass the subcritical envelope (which happens at
a dimensionless time $\tau = 1$ in Fig.~\ref{fig:flow}) and the mass infall rate declines
significantly below its standard value $\sim (1+H_0)a^3/G$, with $H_0$ a
measure of the static magnetic contribution to cloud-core support 
(Li \& Shu 1996;
Allen et al. 2003a,b), the X-wind may begin to blast its way to ever 
wider opening angles
and push on the cloud magnetic fields that thread through the infalling
pseudodisk to reduce the accretion rate further.  The interaction may be
quite complex, as the wind outflow rate itself may be a fraction 
($\sim$ 1/3) of
the mass inflow rate (Shu et al. 1988, 1994, 2000).  Nevertheless, once
outward motions are imparted to a significant portion of the surrounding
cloud envelope, inertia and the ``action at a distance'' that is part
of the effect of magnetic tension, may succeed in blowing away the
infalling pseudodisk, even when the instantaneous wind rate from the 
central star drops to negligible values. Although detailed calculations are
needed to quantify the estimates, which we are in the process of performing,
comparison with the shape of the base of the outflow
cavities in Class I sources suggests that the configuration at $\tau = 2.5$
in Figure~\ref{fig:flow} may be amenable to complete reversal of the 
inflow.  As an approximate guess, therefore, we suppose that 1/2 of the core mass $M_0$
will have dropped into the central regions before an X-wind can 
effectively halt
the infall.  Of this 1/2, only 2/3 has ended up on the star (1/3 has come
out as an X-wind); thus, $M_* = (2/3)(1/2)M_0 = M_0/3$.
The entire description would then lend physical content to the observational
definition that the transition from Class 0 to Class I marks the 
end of the phase of major infall in protostellar evolution (Andr\'e, 
Ward-Thompson, \& Barsony 1993).  

In summary, the answer to the question posed by the title of our
paper is a qualified ``yes'' --  yes, magnetic
levitation and suspension may
help to define the masses of forming stars -- but only in a dynamical
context that involves rapid rotation and protostellar outflows.  Knowing
the relationship between stellar mass to core mass for any single region
then reduces the problem of the stellar IMF to an explanation for the mass
distribution of the cores of molecular clouds.

\acknowledgments{We thank our many colleagues and coworkers, too many 
to mention individually here,
for discussions throughout the years that have shaped our ideas on
star formation and the origin of stellar masses.  After more than five
decades, the problem continues to fascinate and confound.
In Taiwan this work is supported by grants from the National Science
Council (NSC92-2112-M-007-051) and Academia Sinica (to the Theoretical Institute for
Advanced Research in Astrophysics).  In the United States it is supported by 
grants from the National Science Foundation and NASA.}

\bigskip\goodbreak
\centerline
{\bf Appendix A: Comments on the Calculation of $K_0(\eta)$}
\bigskip

The (force) integral-function $K_0(\eta)$ is related to a simpler
(potential) integral-function:
\begin{equation}
H(\xi) \equiv {1\over 2\pi}\oint {d\varphi \over 
\sqrt{1+\xi^2-2\xi\cos\varphi}}. 
\label{e46}
\end{equation}
It is trivial to show that
$-\xi^2 H^\prime (\xi)$ is equal to $K_0(\eta)$
if we set $\xi = 1/\eta$.  On the other hand, by making use of the even
properties of the integrand and transforming $\psi \equiv \varphi/2$,
we may write $H(\xi)$ itself in terms of the first elliptic
integral $(\pi/2)E(\kappa)$:
\begin{equation}
H(\xi) = {1\over 1+\xi}E(\kappa), 
\label{e47a}
\end{equation}
where
\begin{equation}
E(\kappa) \equiv {2\over \pi} \int_0^{\pi/2}{d\psi \over
\sqrt{1-\kappa \cos^2\psi}}, \qquad \kappa \equiv {4\xi\over (1+\xi)^2}.
\label{e47b}
\end{equation}
Unlike the semi-infinite ranges of $\xi$ or $\eta$, $\kappa$ is bounded 
to lie within 0 and 1.  This makes $E(\kappa)$ easy to tabulate and/or 
to approximate by empirical fitting formulae with high numerical accuracy
(See Chapter 17 of Abramowitz and Stegun 1972).  Note that
$E(\kappa)$ has a logarithmic singularity at $\kappa =1$, i.e., 
the potential associated with a unit ring of matter at dimensionless 
radius $r$ is weakly singular at the position of the ring $\varpi=r$.  
The associated radial 
acceleration $K_0$ is large, but changes sign, as one crosses the ring.

\bigskip
\centerline
{\bf Appendix B: Asymptotics of Flat Final State}
\bigskip
We wish to solve the governing equation (\ref{e36a}) for $x\gg 1$ where
$\epsilon \ll 1$.  
For large $x$, the two most dangerous terms are
$x$ times 1 and $-\surd 2$ that appear on the right-hand side of
equation (\ref{e36a}).  We expect $\epsilon(x)$ to be very small
at large $x$, in such a way that the left-hand
side of equation (\ref{e36a}) cannot be expected to balance either of
these terms [otherwise, $\epsilon (x)$ would 
behave either as $-\ln x$ or as $-1/(x\surd 2)$ at large $x$,
both of which violate the expectation that $\epsilon$ is
everywhere positive when matter moves inward in gravitational collapse].
The only way that the two troublesome terms can be eliminated is if
\begin{equation}
{\cal J}(x) \rightarrow -1 \qquad {\rm and}\qquad {\cal I}(x)\rightarrow
-\surd 2 \qquad {\rm as} \qquad x \rightarrow \infty. 
\label{e48}
\end{equation}
These properties are indeed fulfilled by the numerical solution. 

The mathematical reason that the two constraints (\ref{e48}) are fulfilled
individually, rather than in sum, is simple.  Consider a value of
$x$ so large that all of the contributions to the integrals
defining ${\cal I}(x)$ and ${\cal J}(x)$ come from values of
$y \ll x$ [because $\epsilon(y)$ becomes negligible for $y$
comparable to or much larger than $x$].  For $y \ll x$, $K_0(y/x)$ is
well approximated by unity, and equation (\ref{e36b}) and (\ref{e36c})
become
\begin{equation}
{\cal I}(\infty) = \int_{0^+}^\infty \epsilon^\prime(y)\, dy 
= -\epsilon(0^+)= -\surd 2,
\label{e49}
\end{equation}
\begin{equation}
{\cal J}(\infty) = \int_{0^+}^\infty {d\over dy}\left[
y\epsilon(y)+{1\over 2}\epsilon^2(y)\right] \, 
dy = -{1\over 2}\epsilon^2(0^+) = -1.
\label{e50}
\end{equation}
Thus, both constraints of equation (\ref{e48}) are automatically 
fulfilled if our solution satisfies the inner BC, $\epsilon(0^+)
=\surd 2$, and $\epsilon(y)$ goes to zero at large
$y$ faster than $y^{-1}$.  Since we anticipate that $\epsilon(y)$ is an 
analytic function of $y$ at infinity, we suppose it to have a Laurent 
series expansion,
\begin{equation}
\epsilon(y) = {A_2\over y^2}+{A_3\over y^3} + \dots . 
\label{e51}
\end{equation}
By numerical solution of the integro-differential equation (\ref{e36a}),
we determine the coefficient $A_2$ to be close to 1.

\pagebreak

\begin{figure}
  \begin{center}
    \plotone{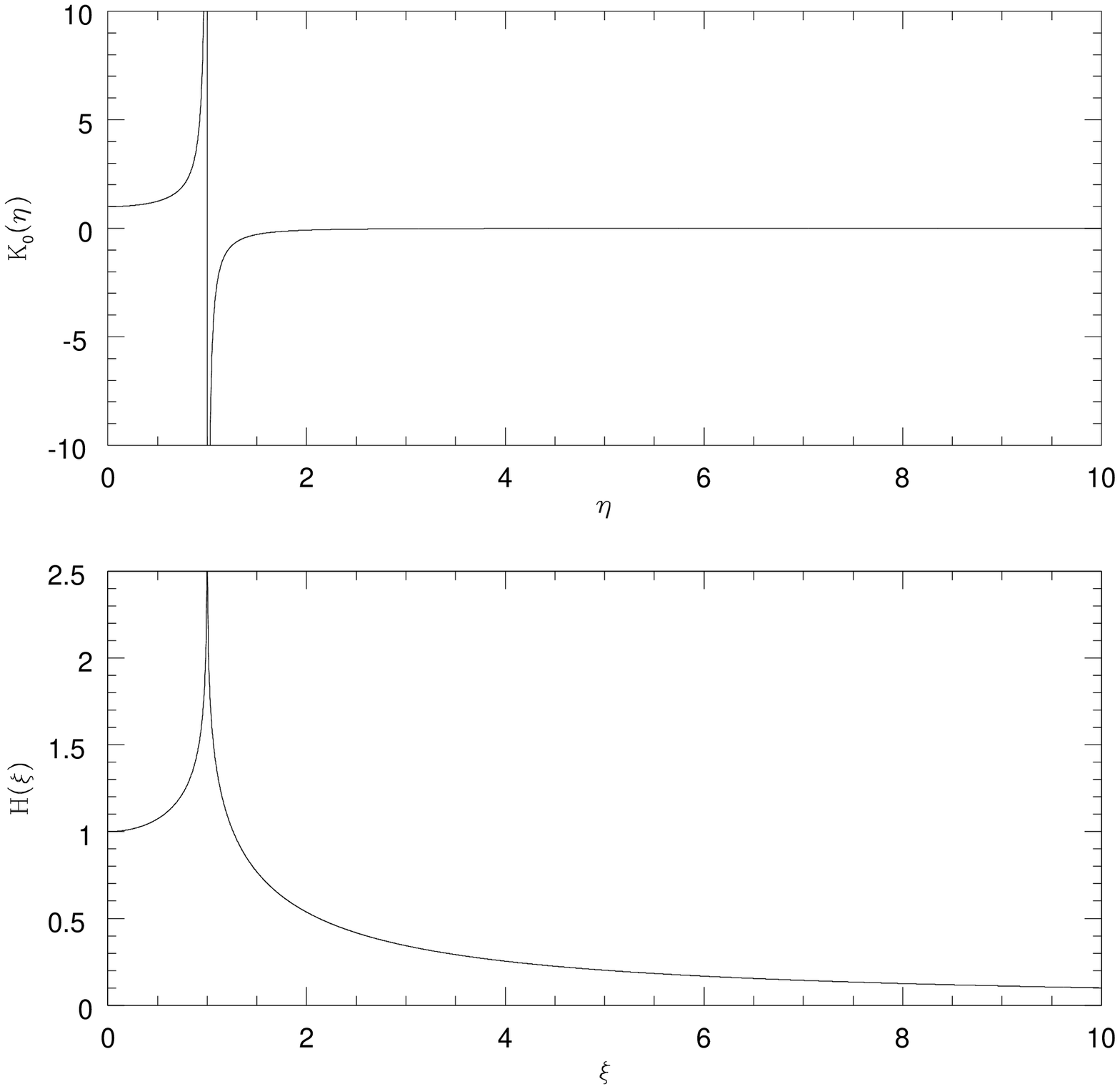}
\caption{The normalized gravity kernel $K_0(\eta)$ and its associated 
``potential'' function $H(\xi)$ defined in Appendix A, with $\xi=1/
\eta$. Notice that both 
functions approach 1 at the origin and zero at the infinity. }
\label{fig:specialfunctions}
  \end{center}
\end{figure}

\pagebreak

\begin{figure}
  \begin{center}
    \plotone{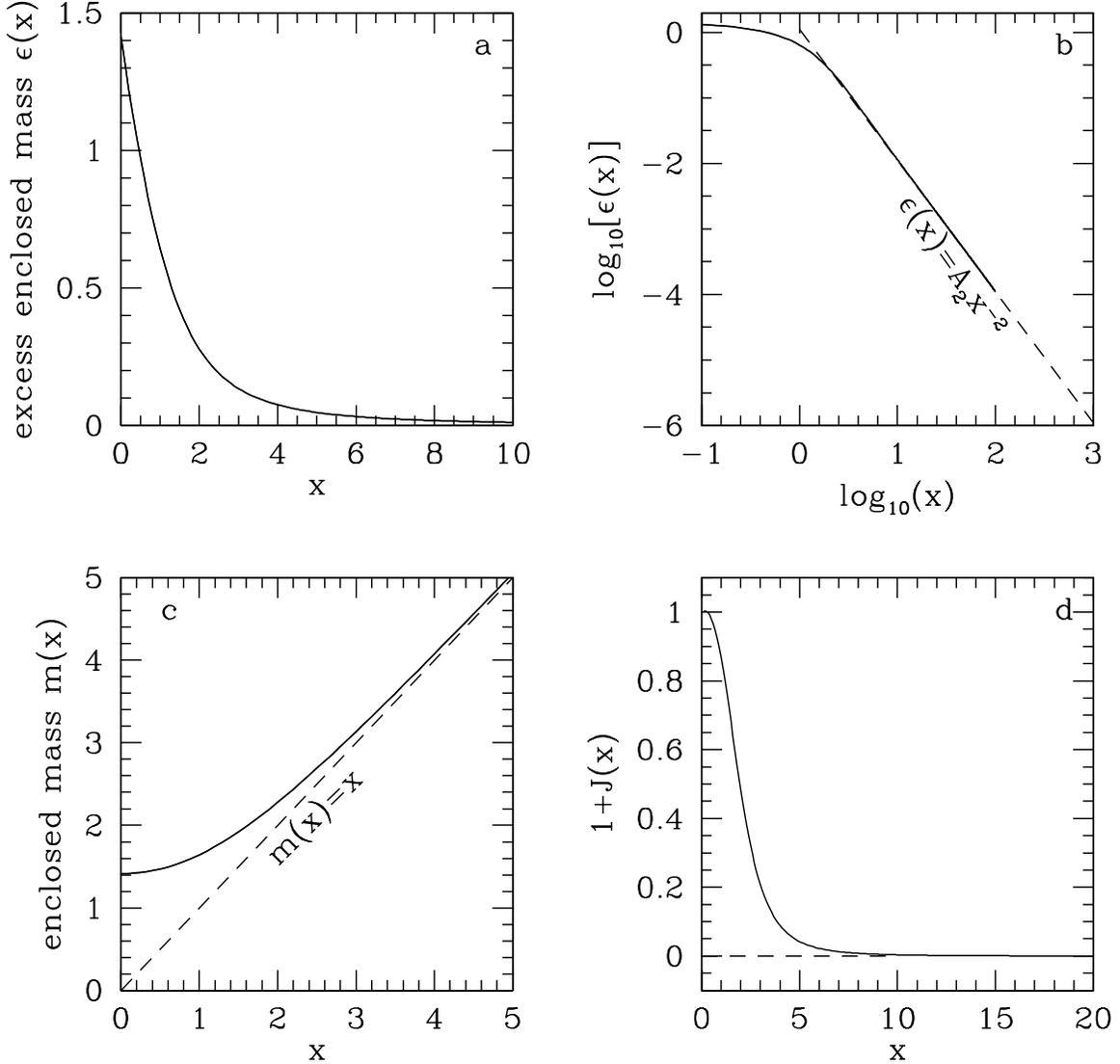}
\caption{Properties of the flat, final state. Shown in panels are 
(a) linear-linear plot of the numerically-determined eigenfunction 
$\epsilon(x)$, which denotes the difference in the mass enclosed 
within a given radius $x$ between the final and initial state; 
(b) log-log plot 
of $\epsilon(x)$, showing the power-law decay of the eigenfunction 
at large radii (the coefficient of the power-law plotted in dashed 
line is $A_2=1.1$); (c) distribution of the enclosed mass $m(x)$ 
in the final state (solid line) compared with that in the initial 
state (dashed line); and (d) the combination $1+{\cal J}(x)$, which 
is proportional to the total magnetic force due to levitation and
suspension.}
\label{fig:eigenfunction}
  \end{center}
\end{figure}
\pagebreak

\begin{figure}
  \begin{center}
    \plotone{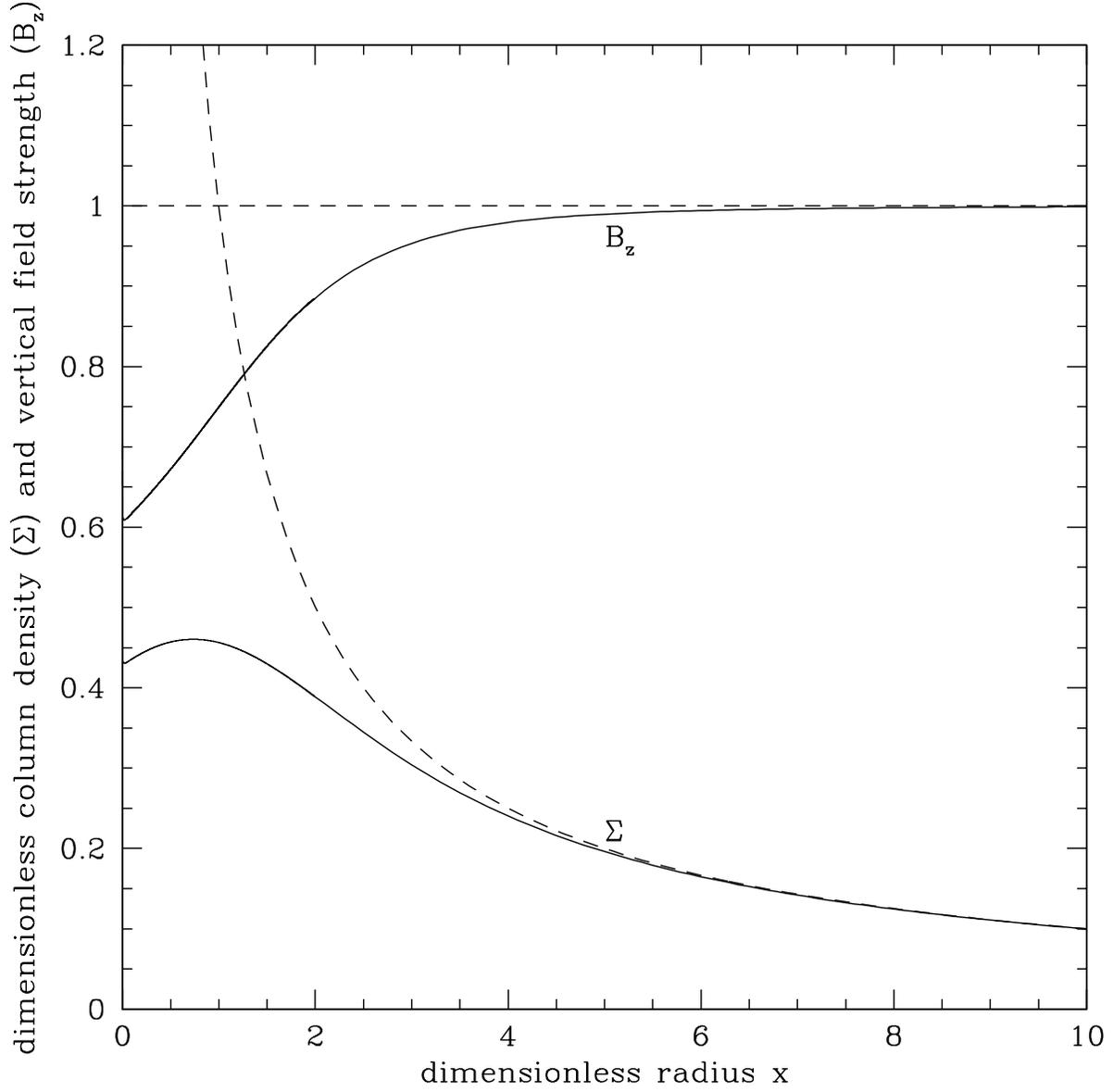}
\caption{Distributions of the (dimensionless) column density and the 
strength of the vertical magnetic field on the flat pseudodisk in
the final state (solid lines). Notice the reduction in both 
the column density and field strength at small radii compared with
the initial distributions (dashed lines). }
\label{fig:sigmaB}
  \end{center}
\end{figure}
\pagebreak

\begin{figure}
  \begin{center}
    \plotone{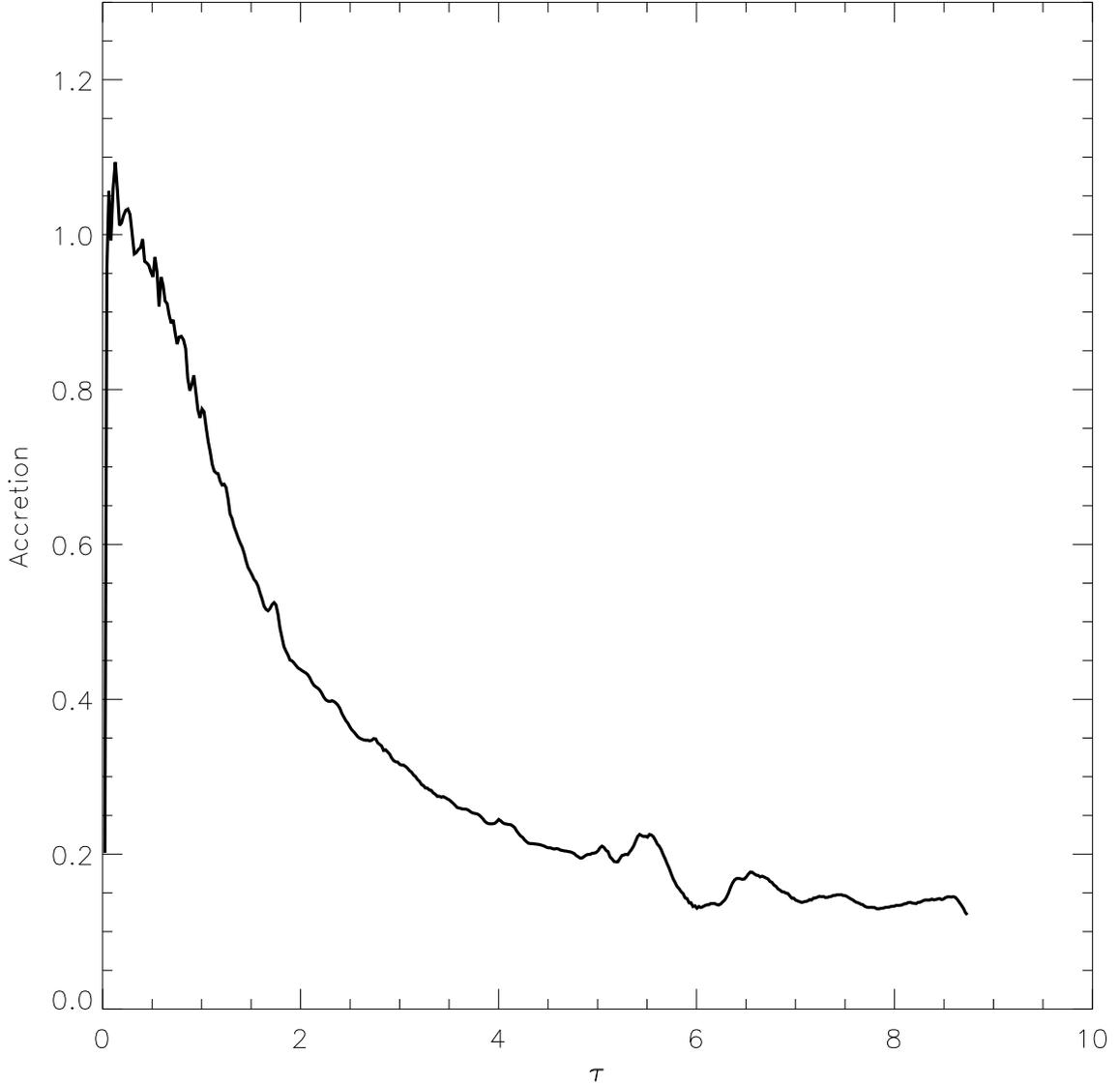}
\caption{The dimensionless mass-accretion rate $\dot m \equiv
G{\dot M}/a^3$ into the central cell
as a function of dimensionless time $\tau \equiv at/r_0$.  The
spike at small times is an artifact of the starting conditions;
$\dot m$ quickly drops to the value 0.975
expected from the unmagnetized SIS analysis (Shu 1977).  Because of
the non-isopedic magnetization of the present calculation, however,
$\dot m$ continues to decline with $\tau$ as the infall encompasses
ever more subcritical envelope.  At $\tau = 5.4$, when the central mass
$m \equiv M/M_0$ is 56\% of the expected final value of $\surd 2$,
growing oscillations of the pseudodisk
caused by numerical difficulties akin to those described in Allen et al.
(2003a,b) lead to increasingly inaccurate representations of the 
regions closest to the central star (see Fig. 6).}
\label{fig:Mdot}
  \end{center}
\end{figure}
\pagebreak

\begin{figure}
  \begin{center}
    \leavevmode
	\epsfxsize=.4\textwidth\epsfbox{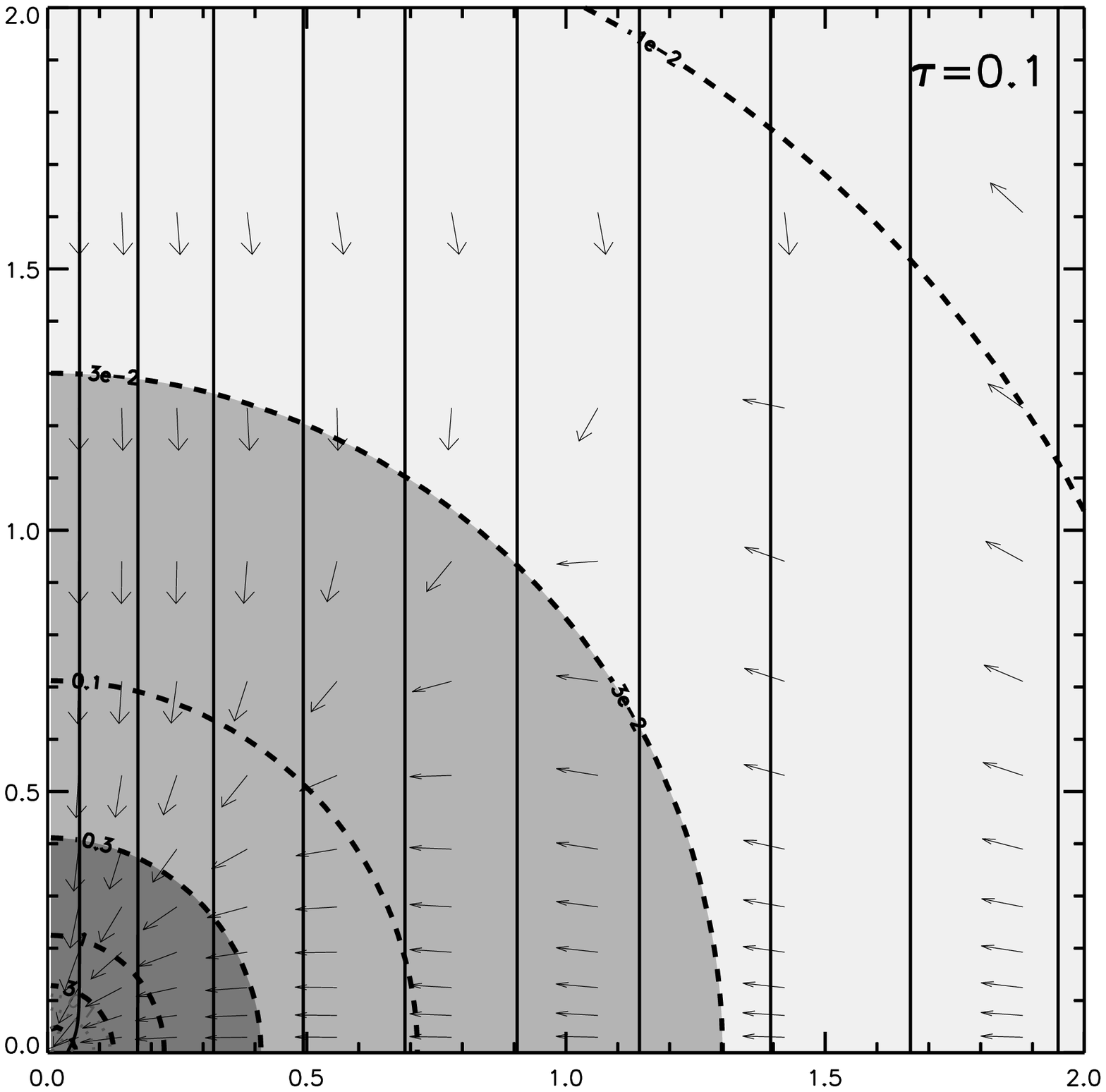}
	\epsfxsize=.4\textwidth\epsfbox{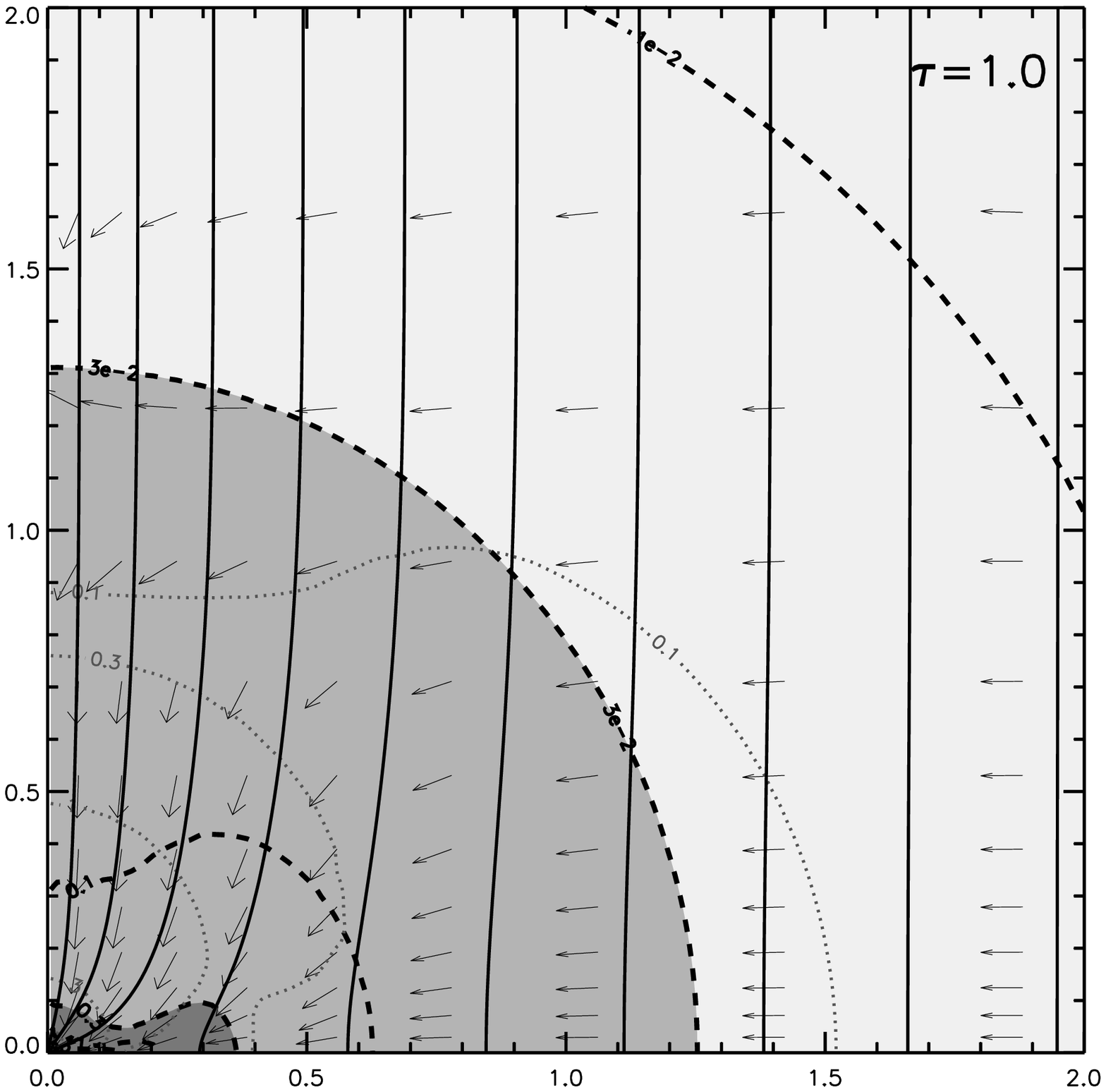}
	\epsfxsize=.4\textwidth\epsfbox{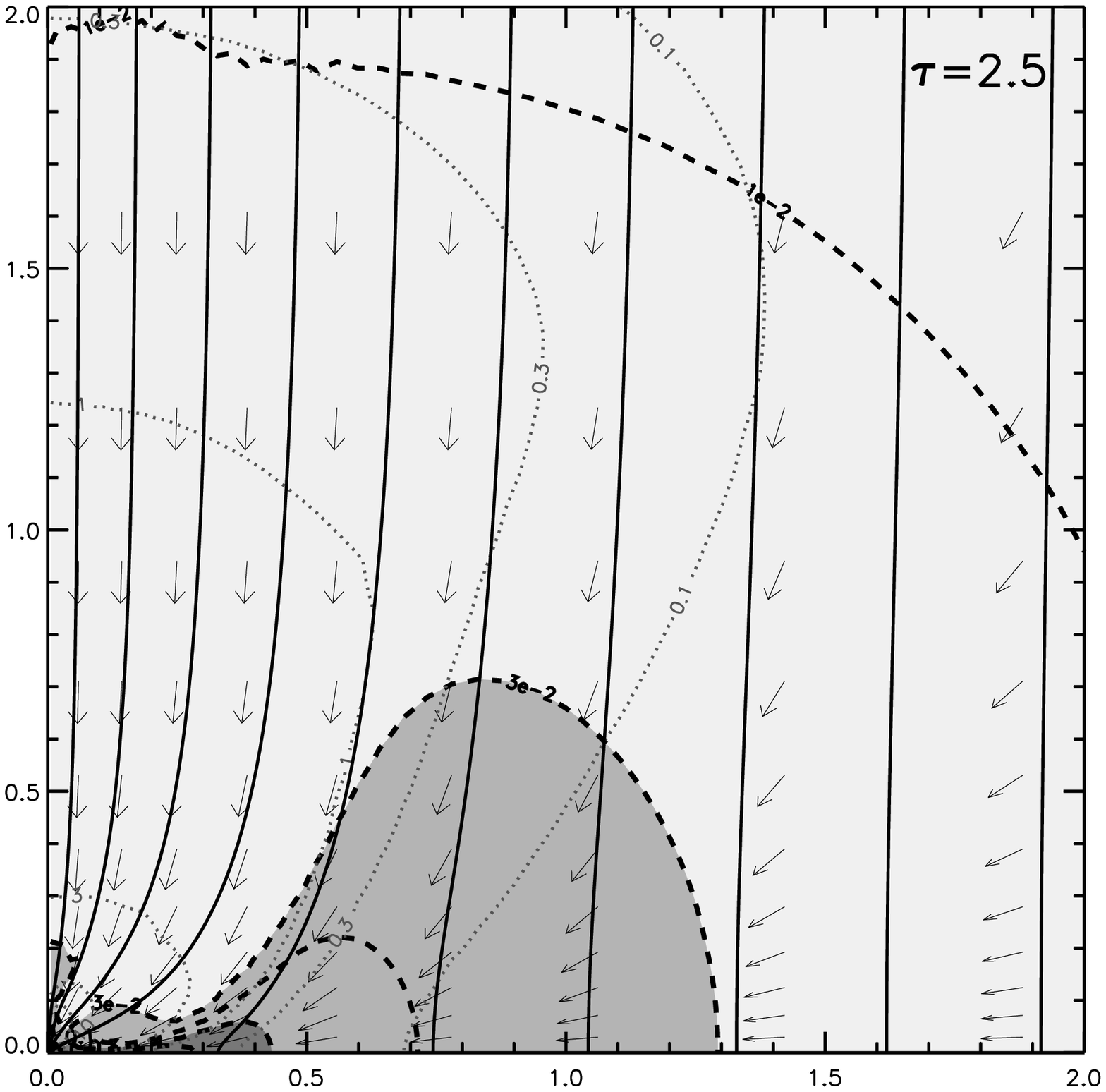}
	\epsfxsize=.4\textwidth\epsfbox{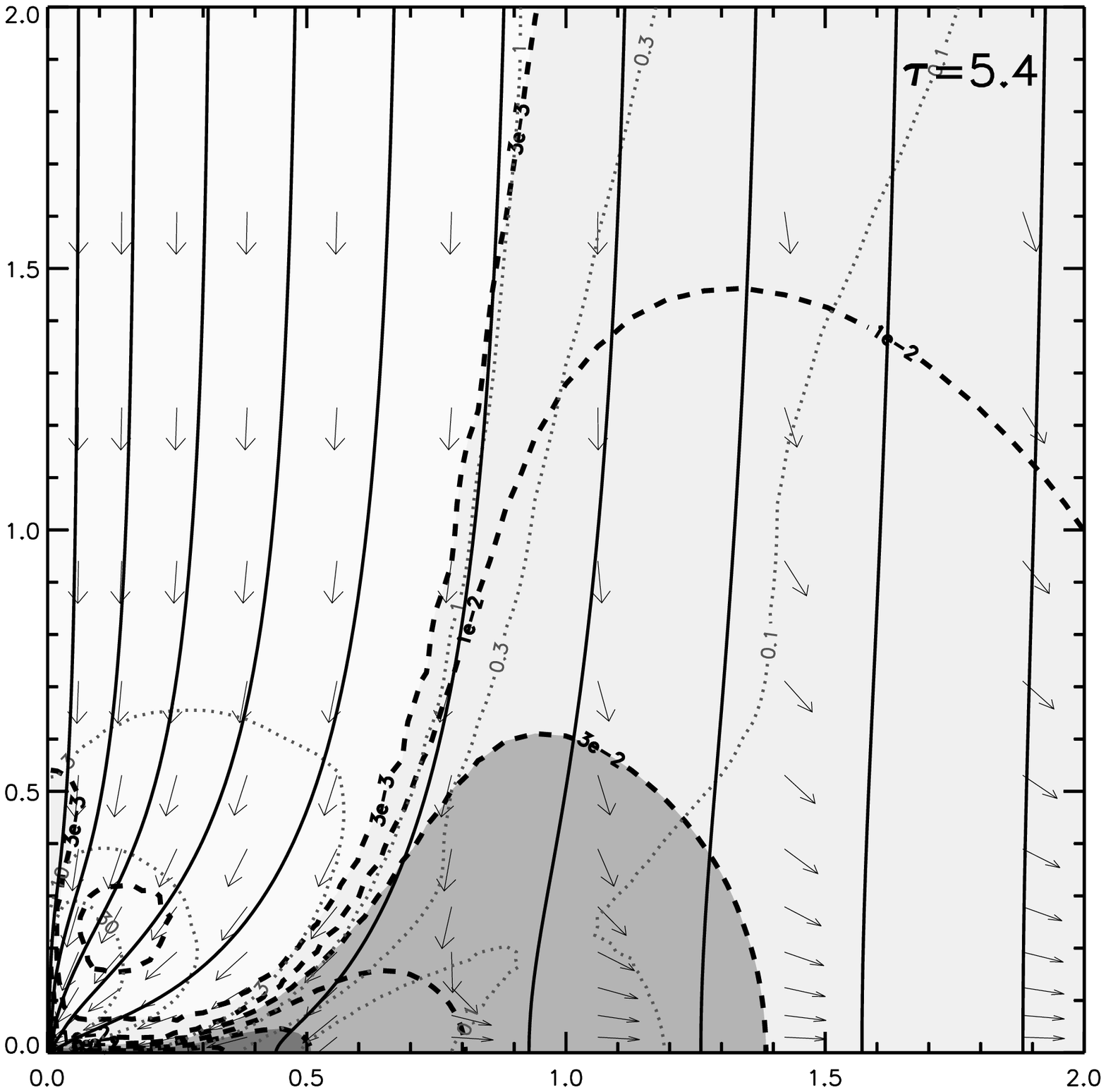}
\caption{The infall flow direction (unit arrows)
and magnetic field lines (solid curves) in the meridional plane
at a dimensionless times $\tau=0.1, 1.0. 2.5, 5.4$.
Distances (axes), isodensity contours (heavy dashed curves),
and isovelociy contours (light dotted curves)
are plotted with $r_0$, $M_0/r_0^3$, and $a$ as the units of time,
density, and speed.  We have not plotted iso-field-strength contours
because, except for the protostar, the entire region has magnetic
field strength between the $0.3 \, B_0$ and $B_0$ values that might
have been used for contouring.
At $\tau = 0.1$, the flow is only slightly perturbed from
what we may expect for the collapse of the unmagnetized SIS.
At $\tau = 1$, an infalling pseudodisk is
apparent at small radii along the midplane.
At large distances from the origin,
a motion occurs primarily toward the magnetic axis,
which carries field lines almost cylindrically to replace those that have been
pinched inward by the point source at the center.
At $\tau = 2.5$, the inward bunching of field lines
has grown sufficiently to prevent much further cylindrical
concentration of field lines, and the gas flow becomes more nearly along field
lines.  As a consequence, a near-vacuum region forms close
to the magnetic axis as most of the material above the midplane
drains onto the pseudodisk.  At $\tau = 5.4$, the
pseudodisk is very flat in the central
regions, $x\equiv \varpi/r_0 \lsim 0.5$, but it flares
into an expected toroid-like structure
farther out where the gravity of the central mass point no longer
dominates the self-gravity and pressure forces of the envelope.
Within the pseudodisk, there is still some radial flow into the central
mass point (see Fig. [\ref{fig:Mdot}]), 
which is still only 56\% of the value $\surd 2$ times $M_0$
expected from the axisymmetric 2-D eigenvalue analysis.}
\label{fig:flow}
  \end{center}
\end{figure}
\pagebreak

\begin{figure}
  \begin{center}
    \plotone{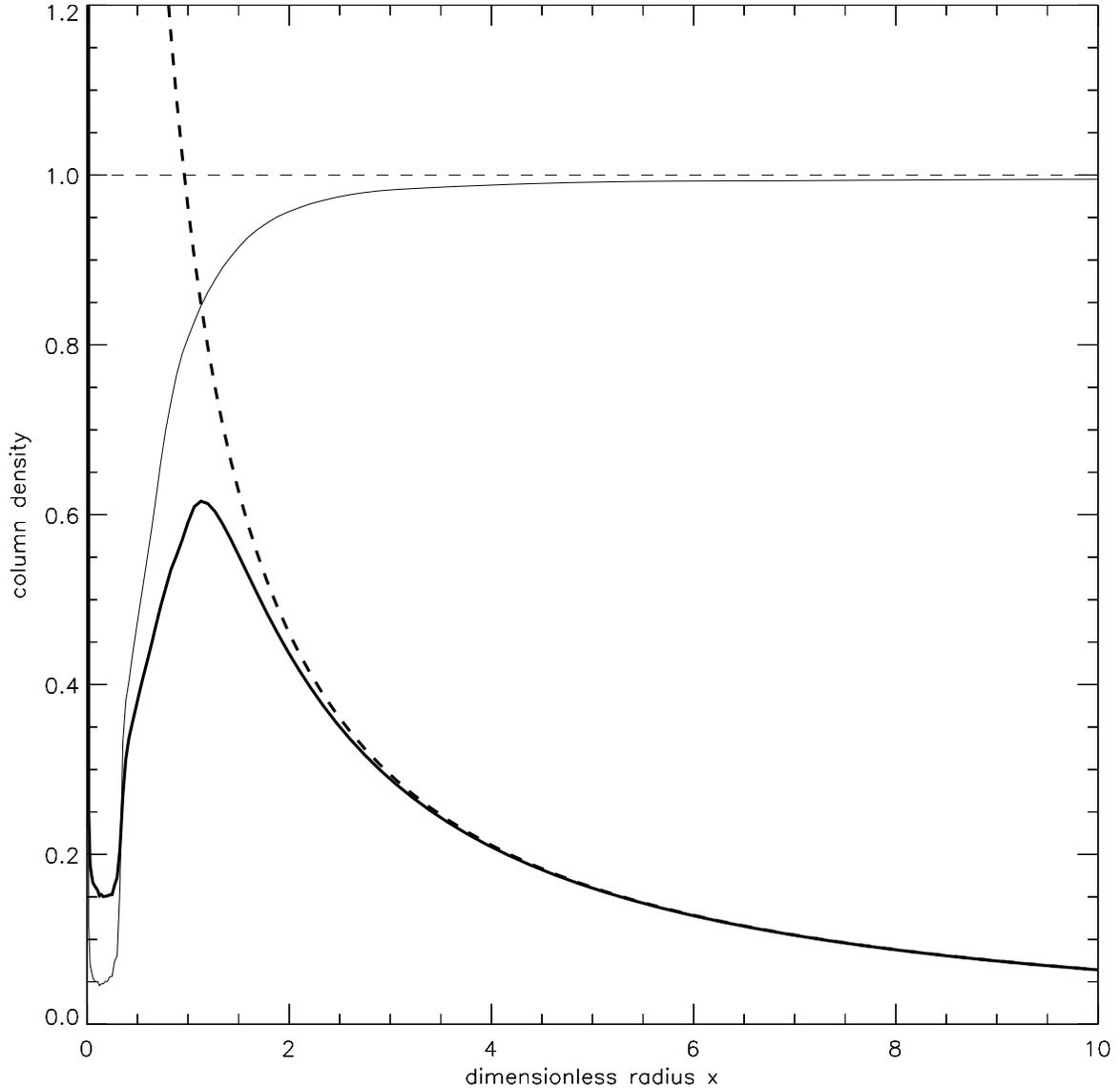}
\caption{Distributions of the (dimensionless) column density and the
strength of the vertical magnetic field at the midplane 
for the 3-D time-dependent (axisymmetric) calculations 
at a dimensionless time $\tau=8.5$ (solid lines) in the
same run as Figures 4 and 5. 
Notice the reduction in both 
the column density and field strength at small radii compared with
the initial distributions (dashed lines). Close to
the protostar, numerical difficulties are being encountered with
field freezing because the curves for (normalized)
$B_z$ and $\Sigma$ should not cross
for the subcritical material outside of the stellar core.  Apart
from the numerical diffusion associated with the mediation of
accretion shocks, this figure compares well with the 2-D
(axisymmetric) final-state results of Figure 3.}
\label{fig:sigmaB3}
  \end{center}
\end{figure}

\begin{figure}
  \begin{center}
   \plotone{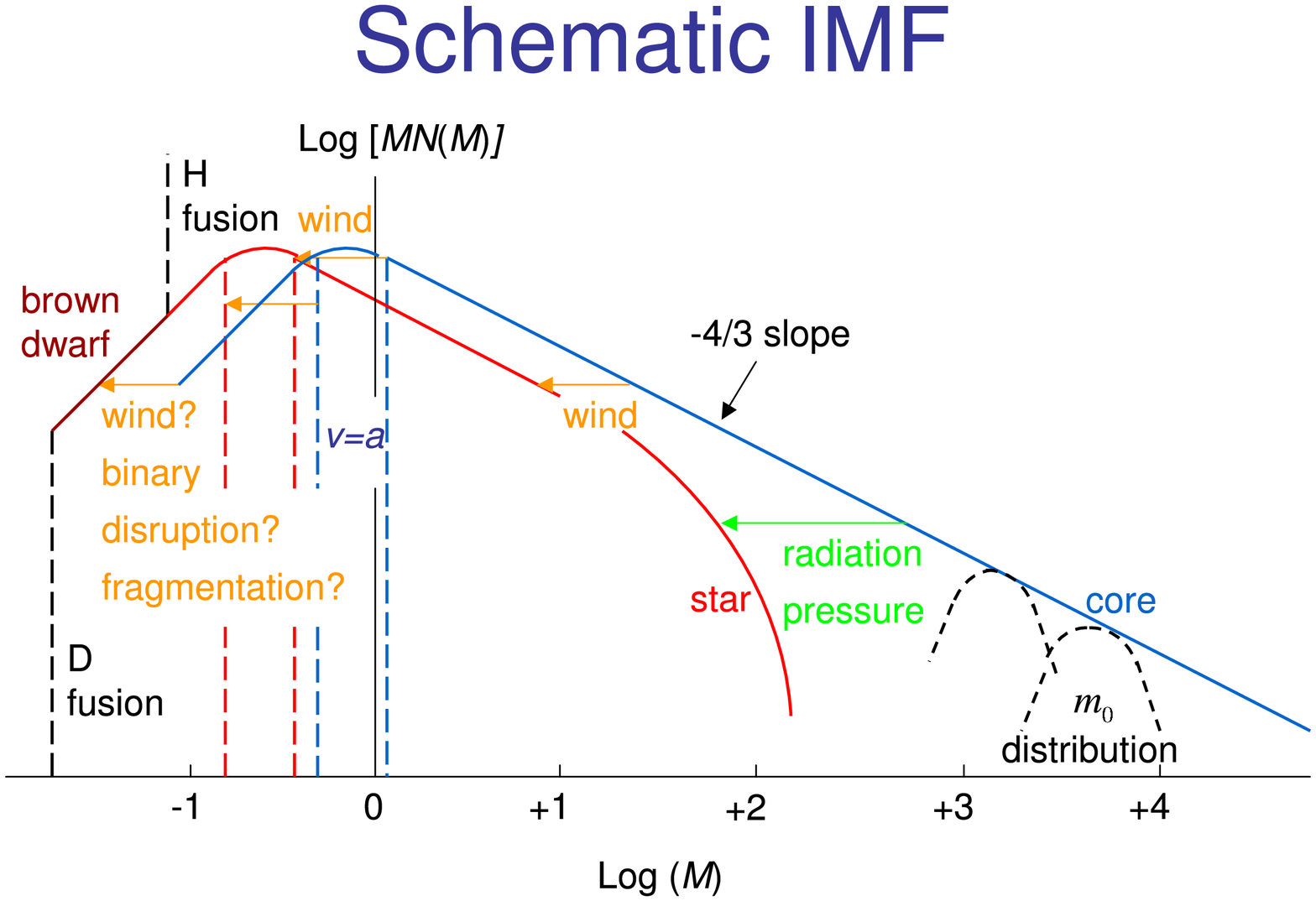}
\caption{The distribution $M{\cal N}(M)$ plotted schematically
for cloud cores and young stars against mass $M$ in a log-log format.
The unit of mass is solar masses, and the vertical scale is arbitrary.
Cloud cores of given $v^4/G^{3/2}B_0$ have a statistical distribution
of masses as indicated by the two bell-functions at large core masses,
because of the variation of the coefficient $m_0$.  The convolution of a
bell distribution with a $-4/3$ power-law distribution
of $v^4/G^{3/2}B_0$ produces
the solid curve labeled by ``core.''  This distribution reaches a natural peak
when $v$ equals the thermal sound speed $a$, which itself has a range of values
indicated schematically by the two vertical dashed lines.
A further extension toward
``subthermal values'' (beyond that expected from the distribution of $m_0$)
may result if fragmentation of thermal cores can occur during the subsequent
collapse, or if binary disruption occurs to unbind substellar companions from
normal stars.  Without fragmentation or binary disruption,
the core mass is reduced
to the star mass by the arrows
depicting the effects of YSO winds and radiation pressure
acting on dust grains during the gravitational collapse of any core.
The hydrogen- and deuterium-burning limits that
separate stars from brown dwarfs and brown dwarfs from (free-floating)
planets are indicated.  No discontinuity at the H-burning limit occurs,
because pre-main-sequence stars are observed to lie in the Hertzsprung-Russell
diagram above the line (Zero-age-main sequence) where they would start
burning hydrogen; thus hydrogen-burning could not have played a role in
the determination of their initial masses.  On the other hand, T Tauri stars
do exhibit a ``birthline'' (Stahler 1983), which has been associated with
deuterium burning, and deuterium burning could play a role in the shape of
the lower-mass end of the stellar IMF (Shu \& Terebey 1984).}
  \end{center}
\end{figure}

\end{document}